\documentclass[11pt,letter,authoryear]{elsarticle}
\usepackage{amsfonts}

\usepackage{amsmath,amssymb,amsthm}
\usepackage{mathtools}
\usepackage{datetime}
\usepackage{enumerate}
\usepackage{xcolor,graphicx}
\usepackage{subfigure}
\usepackage{xfrac}
\usepackage{tabularx}
\usepackage{booktabs}
\usepackage{aligned-overset}
\usepackage{threeparttable}
\usepackage{setspace}
\usepackage{float}
\usepackage{wrapfig}
\usepackage{natbib}
\usepackage{bibentry}
\usepackage{geometry}

\newcommand{\size}{.45}
\newtheorem{thm}{Theorem}
\newtheorem{prop}[thm]{Proposition}
\newtheorem{lem}[thm]{Lemma}
\newtheorem{cor}[thm]{Corollary}

\newdefinition{rem}{Remark}
\newproof{pf}{Proof}
\newproof{pop}{Proof of Proposition \ref{}}
\graphicspath{{../Figures/}}

\begin{document}

\begin{frontmatter}
	
\title{Closed-form portfolio optimization under GARCH models. \tnoteref{t1}}
%\tnotetext[t1]{\color{red}{The authors thank \dots}}

\author{Marcos Escobar-Anel\corref{cor1}}
\ead{marcos.escobar@uwo.ca}
\address{Department of Statistical and Actuarial Sciences, University of Western Ontario, London, ON, Canada, N6A5B7}
\author{Maximilian Gollart}
\ead{maximilian.gollart@tum.de}
\author{Rudi Zagst}
\ead{zagst@tum.de}
\address{Department of Mathematics, Technical University of Munich, Munich, Germany}
\cortext[cor1]{Corresponding author}

\begin{abstract}
This paper develops the first closed-form optimal portfolio allocation formula for a spot asset whose variance follows a GARCH(1,1) process. We consider an investor with constant relative risk
aversion (CRRA) utility who wants to maximize the expected utility from terminal wealth under a \cite{Heston.2000} GARCH (HN-GARCH) model. We obtain closed formulas for the optimal investment strategy, the value function and the optimal terminal wealth. We find the optimal strategy is independent of the development of the risky asset, and the solution converges to that of a continuous-time Heston stochastic volatility model (\cite{Kraft*.2005}), albeit under additional conditions. For a daily trading scenario, the optimal solutions are quite robust to variations in the parameters, while the numerical wealth equivalent loss (WEL) analysis shows good performance of the Heston solution, with a quite inferior performance of the Merton solution.
		

\end{abstract}

\end{frontmatter}

\onehalfspacing
\newpage

\section{Introduction}

The topic of portfolio selection is one of the oldest and still one of the
most discussed research areas in financial economics. \cite{Markowitz.1952}
was a pioneer using mathematical modeling to study this problem. He
presented a framework to optimize portfolios in a mean-variance one-period
setting. Even thought this is probably the most influential portfolio
framework today, at the time only little attention was paid to his work. By
only accounting for one period, he made the assumption that either investors
do not adjust their investment decisions over time as new information
arrives or that they only care about short horizons. In the 60s, \cite%
{Mossin.1968}, \cite{Samuelson.1969} and \cite{RobertC.Merton.1996}
considered a multi-period portfolio problem where, instead of optimizing a
mean-variance trade-off, they maximized expected utility, i.e. Expected
Utility Theory (EUT). \cite{Mossin.1968} was the first to document a dynamic
programming approach to optimize expected utility from terminal wealth. He
chose a discrete-time model for the evolution of wealth using i.i.d.
returns of arbitrary distribution with one risky and one risk-free asset.
\cite{Samuelson.1969} expanded this problem by introducing consumption. A
shift toward continuous-time models, in the seminal work of \cite%
{RobertC.Merton.1996}, permitted closed-form expressions for optimal
consumption and asset allocation under a CRRA utility and a geometric
Brownian motion (GBM) process for the asset price.

The complexity of financial time series has steadily increased in the last
50 years. This fact is supported by the wide range of stylized facts
detected on stock returns, see \cite{Cont.2001} for an overview.
Consequently, plenty of progress has been made in the area of dynamic
process modeling, see for instance \cite{bauwens2006multivariate} for a
survey of generalized autoregressive conditional heteroskedasticity (GARCH)
models. The feasibility of closed-form solutions in
continuous-time portfolio problems has given an advantage to the
continuous-time stream of research, leading to analytical solutions for
extensions of the GBM model. Two representative examples are \cite%
{Kraft*.2005} with the stochastic volatility model of \cite{Heston.1993},
and \cite{liu2003dynamic} adding jumps with stochastic intensities. In
reality, advanced continuous-time models are challenging from an
estimation/calibration perspective. This is mainly due to the presence of
unobservable hidden processes which hurts the stability and efficiency of
estimation methods.

On the other hand, discrete-time models are much more convenient to estimate
with available data, and are more realistic in terms of time frequency for
investor decisions. Nonetheless, multi-period portfolio analysis has seen
limited action since the intensive work of the 60s. One could speculate that this
is due to the lack of closed-form solutions for realistic models. One stream
of the literature has proposed numerical methods for the dynamic portfolio
optimization problem at hand. For instance, \cite{Soyer.2006} presents a
Monte Carlo (MC) based approach where the risky assets follow a multivariate
GARCH model numerically. \cite{brandt2005simulation} relies on the Bellman
principle, Taylor expansions and Least Squares MC to build an approximation.
While \cite{Quek.2017} uses the martingale method for complete markets to
approximate a solution.

A parallel stream has constructed analytical solutions for less realistic
models. For instance, \cite{Canakoglu.2009} present a solution under a
market model that follows a discrete-time Markov chain for exponential
utility. \cite{Dokuchaev.2010} produces a solution where the market model
allows serial correlation in asset returns, leading to a myopic strategy
i.e. independent of time. \cite{Jurek.2011} consider a vector
autoregressive (VAR) model for the dynamics of prices that allows for
stylized facts such as serial correlation of returns and offers flexibility
in modeling the covariance structure among assets. A very similar paper
considering an exponential utility function was published by \cite%
{Bodnar.2015}. In their model, asset returns are assumed to be partly
determined by a set of predictable state variables e.g. dividend yield, term
spread or another asset return.

While all these approaches are improvements over the pioneering work of \cite{Mossin.1968}
and \cite{Samuelson.1969}, deriving these optimal solutions is either very time consuming, or the
models miss well-known stylized facts observed in asset returns. In particular, closed-form solutions to the EUT portfolio problem
in the context of GARCH models, and therefore capturing volatility
clustering, has not been successfully solved yet. This is the main objective
of the paper. For clarity, the contributions of this paper are listed next:

\begin{enumerate}
\item To the best of our knowledge, we are the first to solve in closed-form a dynamic portfolio optimization
problem for a GARCH model (i.e. the HN-GARCH proposed by \cite{Heston.2000}). In particular we produce formulas for the optimal
investment strategy, the value function, the optimal wealth process and its
conditional moment generating function (m.g.f.), in the context of a CRRA investor.

\item Our approach provides the optimal strategy for any given rebalancing
frequency (e.g. intraday, daily, quarterly, etc), connecting to the well-known closed-form solutions from continuous-time models, i.e. one
rebalancing time as per Merton's GBM model, and continuous rebalancing as
per Heston's model.

\item In particular, we prove the convergence of the optimal HN-GARCH
strategy in rebalancing frequency to the optimal strategy in Heston's model (as per
\cite{Kraft*.2005}). The convergence behavior of our solution is shown numerically to be slightly non-linear for some risk aversion levels.

\item We illustrate the impact of the various GARCH parameters
on the optimal investment strategy, demonstrating the solution is quite robust against deviations
from the true parameter values e.g. inaccurate estimations.

\item The impact of the self-financing approximation is shown to be
negligible in terms of the wealth process and extra cash flows.

\item We study the wealth-equivalent loss (WEL) incurred by an investor who
trades daily as per our model, but uses popular closed-form continuous-time
solutions (e.g. GBM or Heston model) instead of our optimal. The analysis
demonstrates a good performance by Heston, which is only compromised in cases of high levels of
market price of risk. And a quite poor performance of Merton's solution
across most of the parametric space.
\end{enumerate}

The paper is organized as follows: Section \ref{sec:Setting
general} introduces the mathematical setting and lines out our approach to
obtain the closed-form solution. Section \ref{sec:SECOND ORDER OPT} presents
the main results and derives the continuous-time limit of our optimal
strategy. Section \ref{sec:second order num} presents numeric analysis
dealing with the impact of approximating the wealth process, the sensitivity
of our solution towards various parameters, the convergence behavior of our
solution and a comparison to other well-known portfolio strategies. Section %
\ref{sec:CONCLUSION} concludes the paper. Most proofs are provided in the
Appendix or the supplementary online material.

\section{Mathematical setting and outline of the approach. \label%
{sec:Setting general}}

Let ($\Omega$, $\mathcal{F}$, $\mathbb{P}$) be a complete probability space
with filtration $\left\{\mathcal{F}_{t}\right\}_{t\in\left\{0,1,\dots
T\right\}}$. All stochastic processes are defined on this probability space.
In this setting the log of the risky spot asset price $X_{t} = \log S_{t}$
is $\mathcal{F}_{t}$-progressively measurable and follows the Heston-Nandi
GARCH (1,1) model,
\begin{align}
& X_{t}=X_{t-1}+r+\lambda h_{t}+\sqrt{h_{t}}z_{t}, \quad X(0)=x_{0}>0
\label{HN log stock} \\
& h_{t}=\omega +\beta h_{t-1}+\alpha (z_{t-1}-\theta \sqrt{h_{t-1}})^{2}
\label{HN GARCH}
\end{align}%
where $x_{0}$ is non-random, $r$ is the continuously compounded
single-period risk-free rate, $z_{t}$ is a standard normal disturbance and $%
h_{t}$ is the conditional variance of the log return of the asset between $%
t-1$ and $t$ with $\beta +\alpha \theta ^{2}<1$ ensuring stationarity. Assuming variance stationarity, the long-term average of the variance ($h$)
and the conditional covariance between the variance and the log-stock are given by
\begin{align}
\bar{h}&=\dfrac{\alpha +\omega }{1-\beta -\alpha \theta ^{2}}  \label{h} \\
\mathbb{C}ov_{t-1}[h_{t+1}, X_{t}] &= -2\alpha \theta h_{t} .
\label{cov X h}
\end{align}
%\textcolor{red}{We may also put the formulas for the covariance $\mathbb{C}ov_{t-1}[h_{t+1}, X_{t}] = -2\alpha \theta h_{t}$, the long term average variance $h=\dfrac{\alpha +\omega }{1-\beta -\alpha \theta ^{2}}$ and the stationarity condition as a remark here so that we can refer to them later e.g. for the parameter sensitivity analysis.}

\begin{lem}
\label{lem:mult period exp h}  The multi-period expectation of (\ref{HN
GARCH}) can be written as
\begin{align*}
\mathbb{E}[h_{t}] = \left(\alpha + \omega \right) \left(\frac{1-\left(\beta
+\alpha \theta ^{2}\right)^{t}}{1-\left(\beta +\alpha \theta ^{2}\right)}%
\right) + (\beta + \alpha \theta^{2})^{t}h_{0}, \quad t\in \mathbb{N}_{0}.
\end{align*}
\end{lem}

\begin{pf}
	It follows directly from the recursive representation in Equation \ref{HN GARCH}. See the complimentary material. %See \ref{proofSpec 1}.
	
	\hfill $\square$
\end{pf}

The second asset is the risk-free bank account $B_{t} $ bearing the
continuously compounded interest rate $r$ for the time interval from $t$ to $%
t+1$.

The proportion of wealth invested in the risky asset $S_{t}$ at any time $t$
is defined as $\pi_{t}$. The remaining wealth $(1-\pi_{t})$ goes into the
risk-free cash account $B_{t}$. We will work in the space $\mathcal{U}[0,T]$
of admissible strategies $\pi \coloneqq \left\{\pi_{t}\right\}_{t\in\left%
\{0,1,\dots T\right\}}$, satisfying three conditions: $\pi_{t}$ is $\mathcal{F}_{t}$%
-progressively measurable, wealth is non-negative in $[0,T]$ and the expectation in
(\ref{P(W) 1}) is well-defined. We further assume that our market is
frictionless and that the risky asset pays no dividends.

Next, we construct the portfolio value $V_{t}$ (wealth) using a
self-financing argument. Let $\varphi_{S,t}$ denote the number of stocks and
$\varphi_{B,t}$ the number of units in the cash account at time $t$. The
value of a portfolio mustn't change through re-balancing, i.e. the value must
be the same with and without re-balancing. This implies
\begin{align}
V_{t} &= \varphi_{S,t} S_{t} + \varphi_{B,t} B_{t} = \varphi_{S,t-1} S_{t} +
\varphi_{B,t-1} B_{t},
\end{align}
where $\pi_{t-1}=\frac{\varphi_{t-1}
S_{t-1}}{V_{t-1}} \Leftrightarrow \varphi_{t-1} = \frac{%
\pi_{t-1}V_{t-1}}{S_{t-1}}$. Using $\pi_{t-1} \coloneqq \pi_{S,t-1} = 1 -
\pi_{B,t-1}$, we can rewrite the condition as
\begin{align}
\frac{V_{t}-V_{t-1}}{V_{t-1}} &=\pi _{t-1}\frac{
S_{t}-S_{t-1}}{S_{t-1}}+\left( 1-\pi _{t-1}\right) \frac{B_{t}-B_{t-1}}{%
B_{t-1}}.  \label{SFC}
\end{align}
This is the exact self-financing condition (SFC), which can be simplified to,
\begin{align}
\frac{V_{t}}{V_{t-1}} &= \pi_{t-1} \frac{S_{t}}{S_{t-1}} + \left(1-\pi_{t-1}\right)e^{r}.  \label{exact wealth}
\end{align}

The GARCH modeling of stocks targets log prices rather than returns, e.g. equation (\ref{HN log stock}). Therefore, we aim at modeling log wealth rather than wealth returns. This means we must approximate all returns in the self-financing condition by log prices. This is done via a Taylor expansion of order two presented next.\footnote{Order two ensures a convergence to the continuous-time solution as opposed to order one. Higher order approximations could also be entertained with no impact on the continuous-time limit.}
%Moreover, a returns-based representation could lead to negative wealth without further constraints on strategies. An alternative, which avoids constraints, is to approximate the log-return process via a Taylor expansion.

%Note, to preserve the nice properties of the HN-GARCH model we need to adopt an approximated wealth process which will be presented in Section \ref{sec:Setting}.
Using a Taylor series expansion around $1$ and working with the variance of
the return instead of the squared return as per the continuous time
counterpart, we can approximate the log-return process of $S_{t}$ as
\begin{align}
\log \left(\frac{S_{t}}{S_{t-1}}\right) &\approx \frac{S_{t}-S_{t-1}%
}{S_{t-1}} - \frac{1}{2} \left(\frac{S_{t}-S_{t-1}}{S_{t-1}}%
\right)^{2} \approx \frac{S_{t}-S_{t-1}}{S_{t-1}} - \frac{1}{2}
\mathbb{V}ar\left[\frac{S_{t}-S_{t-1}}{S_{t-1}}\right].
\label{taylor return}
\end{align}
This can be done analogously for $B_{t}$ and $V_{t}$, i.e. $\frac{B_{t}-B_{t-1}}{B_{t-1}} \approx \log \left(\frac{B_{t}}{B_{t-1}}\right) = r$ and $\frac{V_{t}-V_{t-1}}{V_{t-1}}\approx \log \left(\frac{V_{t}}{V_{t-1}}\right)+\frac{1}{2}\mathbb{V}ar_{t}[\frac{V_{t}-V_{t-1}}{V_{t-1}}]=W_{t}-W_{t-1}+\frac{1}{2}\pi _{t-1}^{2} h_{t}$ with $W_t=\log(V_t)$. Using these approximations on (\ref{SFC}) and rearranging terms leads to:
\begin{align}
W_{t}&= W_{t-1}+\pi _{t-1}Y_{t} +\left( \pi _{t-1}-\pi
_{t-1}^{2}\right) \frac{1}{2}h_{t}+\left( 1-\pi _{t-1}\right) r, \quad Y_t\coloneqq X_{t}-X_{t-1}. \label{W new}
\end{align}
From now on, we will work with $W_t$ instead of $V_t$ and equation (\ref{W new}) instead of equation (\ref{exact wealth}), i.e. with the approximation of the self-financing condition. The negligible impact of the approximation will be studied in Subsection \ref{sec:proxy 1}. Similar approximations have been used earlier by \cite{Campbell.1999}, \cite%
{Campbell.2003b} and \cite{Jurek.2011}.

Substituting the Heston-Nandi model for $X_{t}$ we get%
\begin{align}
W_{t}-W_{t-1} &=\pi _{t-1}\lambda h_{t}+\left( \pi _{t-1}-\pi
_{t-1}^{2}\right) \frac{1}{2}h_{t}+\pi _{t-1}\sqrt{h_{t}}z_{t}+r
\label{log wealth 1} \\
h_{t} &=\omega +\beta h_{t-1}+\alpha (z_{t-1}-\theta \sqrt{h_{t-1}})^{2}.
\notag
\end{align}

We now choose a power utility function of the form $U\left( v\right) =\frac{%
v^{\gamma }}{\gamma }$. This power utility characterizes the investor as
having a constant level of relative risk aversion (CRRA) of $(1-\gamma)$, implying that for a decreasing $\gamma$ risk aversion increases. For our
portfolio optimization problem we will need to put a constraint on the risk
aversion parameter $\gamma$. That is,\footnote{%
This is a necessary condition for our work, the numerical section suggest
our results are still valid for $\gamma < 1, \gamma\neq0$.}
\begin{align}  \label{myAssumption}
\gamma&<0.
\end{align}

The problem of interest is to maximize expected utility from terminal wealth%
\footnote{%
We do not consider consumption for the sake of simplicity.} i.e. to find a
strategy $\pi^{\ast}$ which solves the optimal control problem\footnote{%
The dependence of $V_{T}$ on the investment strategy is omitted from the
notation.}
\begin{align}
\underset{\left\{ \pi _{t}\right\} _{t=0}^{T-1}}{\max }%
\mathbb{E}_{0}\left[ U\left( V_{T}\right) \right] &= \underset{\left\{ \pi _{t}\right\} _{t=0}^{T-1}}{\max }%
\mathbb{E}_{0}\left[ \frac{\exp \left\{ \gamma W_{T}\right\} }{\gamma }%
\right] = \Phi_{0} (w_{0},h_1), \quad w_0=\log(v_0)   \label{P(W) 1}
\end{align}
We use Bellman's principle to solve the problem recursively period by
period starting at time $T$. A more detailed outline of this idea is
presented below. We start by formulating the following stochastic control
problem:

Let $\mathbb{W}\subset\mathbb{R}$ be the set of possible wealth, $\mathbb{A}$
be the set of admissible portfolios, $\mathbb{H}=(0,\infty)$, and $\mathbb{Y}=(0,\infty)$. The
transition function ($\mathbb{T}$) from $\mathbb{W}\times\mathbb{A}\times \mathbb{H}\times\mathbb{Y}$
to $\mathbb{W}$ is given by (\ref{W new}). Thus, we define
\begin{align}
\mathbb{T}(W,a,h, Y) &\coloneqq W+aY +\left(a-a^{2}\right) \frac{1}{2}h+\left( 1-a\right) r, \quad h>0.
\label{Trans}
\end{align}
Further, let the operators $\mathbb{L}$ and $\mathbb{U}$ be well defined for all admissible functions $\Phi$ by.
\begin{align}
\mathbb{L}\Phi(W,a,h) &\coloneqq \mathbb{E}\left[\Phi(\mathbb{T}(W,a,h, Y))\right], \quad
W \in \mathbb{W}, a \in \mathbb{A}, h \in \mathbb{H} \\
\mathbb{U}\Phi(W,h) &\coloneqq \underset{a\in \mathbb{A}}{\max}\mathbb{L}%
\Phi(W,a,h), \quad W\in \mathbb{W}, h \in \mathbb{H}
\end{align}
A function $\Phi$ is called admissible if there exists a set of functions
\begin{align}
\mathbb{M} \subset \left\{\Phi:%
\mathbb{W}\times \mathbb{H}\rightarrow\mathbb{R}: \mathbb{E}\left[|\Phi|\right]<\infty, \Phi
\text{ concave in } \mathbb{W}  \right\}
\end{align}
such that $\mathbb{U}: \mathbb{M}\rightarrow \mathbb{M}$, $%
\Phi_{0}(w_{0},h_1)\in \mathbb{M}$ and that for all $\Phi\in \mathbb{M}$ there exists an $a\in
\mathbb{A}$ such that $a(W,h)$ maximizes $a\mapsto\mathbb{L}\Phi(W,a,h)$ on $%
\mathbb{A}$ for all $W \in\mathbb{W}$ and $h \in \mathbb{H}$.\footnote{%
In general the existence of $\pi_{t}$ is not necessarily given,
however in our application this is always the case. Thus, we also use the $%
max$, instead of the $sup$ operator in the definition of $%
\mathbb{U}$.}

As per the Bellman principle (cf. \cite{Seierstad.2009} Theorem 1-13), optimizing recursively step by step yields the
optimal strategy $\pi^{\ast}$ for problem (\ref{P(W) 1}). Thus, we can
solve the problem via the value iteration
\begin{align}
\Phi_{t} (W_{t},h_{t+1}) &= \mathbb{U}\Phi_{t+1} (W_{t},h_{t+1}), \quad t=0,\dots,T-1.
\label{EUT_rec}
\end{align}
This notation and equation (\ref{P(W) 1}) require the terminal condition
\begin{align}
\Phi_{T}(W_{T},h_{T+1}) &= \Phi_{T}(W_{T})=U\left(W_{T}\right).  \label{Terminal Condition 1}
\end{align}
For a power utility, we first find $\pi _{T-1}^{\ast }(W_{T-1},h_T)$ as the maximizer
\begin{align}
\Phi_{T-1} (W_{T-1},h_T)&=\mathbb{U}\Phi_{T}(W_{T-1},h_T) = \underset{a\in\mathbb{A}}{\max }\mathbb{L}\Phi_{T} (W_{T-1},a,h_T) =\underset{a\in\mathbb{A}}{\max }\mathbb{E}_{T-1}%
\left[ \Phi_T(\mathbb{T}(W_{T-1},a,h_T,Y_T))\right]
\end{align}
then going after $\pi _{T-2}^{\ast }(W_{T-2},h_{T-1})$ in
\begin{align}
\Phi_{T-2} (W_{T-2},h_{T-1})&=\underset{a\in\mathbb{A}}{\max }\mathbb{E}_{T-1}\left[
\Phi_{T-1} (\mathbb{T}(W_{T-2},a,h_{T-1},Y_{T-1}))\right]
\end{align}
and continue recursively till reaching $t=0$.

This type of portfolio choice problem, where
Bellman's optimality principle can be applied is usually called time
consistent. If this is not the case the problem is called time inconsistent
(see for example \cite{Bensoussan.2014}). This can for example happen when
considering stochastic interest rates like in \cite{Wu.2018}.

The
solution we obtain on $W_t$ for problem (\ref{P(W) 1}) is summarized in \textit{Theorem \ref{myTheorem 1}}. The impact of the
approximation will be assessed later in this paper.

\section{Portfolio optimization solution. \label{sec:SECOND ORDER OPT}}

In this section we solve the portfolio optimization problem (\ref{P(W) 1}).
Subsection \ref{sec:results1} presents the main results. Therein, we also
present properties of the wealth process that is generated by the optimal
strategy. The continuous-time limit of the solution can be found in
Subsection \ref{sec:cont time 1}. Subsection \ref{sec:subopt} introduces the
concept of wealth-equivalent loss and shows how it can be calculated in our
setting.

\subsection{Main results \label{sec:results1}}

Assume the model setting described in Section \ref{sec:Setting general}, in
particular equations (\ref{log
wealth 1}) with the expected utility maximization problem in equation (\ref%
{P(W) 1}). Further, assume that\newline
$\mathbb{M}\coloneqq\left\{\Phi: \mathbb{W}\times \mathbb{H}\rightarrow\mathbb{R}:\Phi(W,h)=\frac{1}{\gamma}e^{D + \gamma W + E h}, D \in \mathbb{R}, E \in \mathbb{R} \right\}$.

\begin{thm}
\label{myTheorem 1}  Let $\gamma<0$. Then, for the problem described in
equation (\ref{P(W) 1}) where the log wealth follows the dynamics in (\ref%
{log wealth 1}) and $1-2 \alpha E_{t,T}^{\ast}>0$, we have for any time t:

\begin{itemize}
\item the optimal expected utility from terminal wealth is given by
\begin{align}
\Phi_{t} (W_{t}, h_{t+1})&=\frac{1}{\gamma }\exp \left\{ D_{t,T}+\gamma
W_{t}+E_{t,T}^{\ast }h_{t+1}\right\}  \label{ValueOpt_pert 1}
\end{align}
where
\begin{align}
D_{t,T} &= D_{t+1,T} + E_{t+1,T}^{\ast }\omega+\gamma r -\log \left( \sqrt{%
(1-2\alpha E_{t+1,T}^{\ast })}\right),\qquad t=0,\dots,T-1
\label{OptA_pert1} \\
E_{t,T}^{\ast } &= \left( \beta +\alpha \theta ^{2}\right)E_{t+1,T}^{\ast } +%
\dfrac{\left( \gamma \pi_{t}^{\ast}-2\theta \alpha E_{t+1,T}^{\ast}\right)
^{2}}{2(1-2\alpha E_{t+1,T}^{\ast})} + \gamma \left((\lambda + \frac{1}{2}%
)\pi_{t}^{\ast}-\frac{1}{2}(\pi_{t}^{\ast})^{2}\right),\qquad t=0,\dots,T-1
\label{OptC_pert 1}
\end{align}
with
\begin{align}
D_{T,T} &= E_{T,T}^{\ast } = 0  \label{TC 1}
\end{align}

\item the optimal proportion invested in the risky asset at any time t is
\begin{align}
\pi _{t}^{\ast }&=\frac{\lambda + \frac{1}{2}}{(1-2 \alpha
E_{t+1,T}^{\ast})-\gamma} - \frac{\left(\theta+(\lambda+\frac{1}{2})\right)
2\alpha E_{t+1,T}^{\ast}}{(1-2 \alpha E_{t+1,T}^{\ast}) - \gamma}
\label{OptAllocation_periodt 1}
\end{align}

\item the optimal wealth is
\begin{align}
W^{\ast}_{t} &= w_{0}+ \sum_{k=1}^{t} \left((\lambda +\frac{1}{2})\pi
_{t-k}^{\ast}-\frac{1}{2}(\pi_{t-k}^{\ast})^{2}\right) h_{t+1-k} +
\sum_{k=1}^{t} \left(\pi_{t-k}^{\ast} \sqrt{h_{t+1-k}}z_{t+1-k}\right) +rt
\label{W_opt_1}
\end{align}
\end{itemize}
\end{thm}

\begin{pf}
	See \ref{proofSpec 1}.
	
	\hfill $\square$
\end{pf}
We note that $\pi_{t}^{\ast}$ is a deterministic process as it does neither
depend on the movements of the investors wealth, nor on the variance.

%Furthermore, observe that $D_{t,T}$ can also be written as
%\begin{align}
%D_{t,T} = (T-t)\gamma r + \sum_{i=0}^{T-t-1}\left(E_{T-i,T}^{\ast}\omega - \log\left(\sqrt{1-2\alpha E_{T-i,T}^{\ast}}\right) \right).
%\end{align}
%{\color{red}Do you think we should keep this for the paper? If so, should we just state this or show the small induction proof from chapter 2 in the thesis?}

In the proof of Theorem \ref{myTheorem 1} we use the assumption of a
negative $\gamma$ and the condition $1-2 \alpha E_{t,T}^{\ast}>0$. However,
they are not restrictive for the parameter set considered in the numeric
section. Numerical experiments suggest that for $\gamma<0$ usually $%
E_{t,T}^{\ast}\leq0$ with $E_{t-1,T}^{\ast}\leq E_{t,T}^{\ast}, \forall
t=0,\dots,T $ which implies that the condition is satisfied. From a
theoretical perspective one can show that, if $E_{t,T}^{\ast}$ is
monotonously increasing in $t$ then the condition is always fulfilled for $%
\gamma<0$. This result is summarized in the following lemma.

\begin{lem}
\label{lemma well defined 1}  Let $E_{t,T}^{\ast}$ be monotonously
increasing in $t$ and let $\gamma<0$, then the condition $1-2 \alpha
E_{t,T}^{\ast}>0$ is always fulfilled.
\end{lem}

\begin{pf}
	We know that $E_{T,T}^{\ast}=0$. If $E_{t,T}^{\ast}$ is monotonously increasing in $t$, this implies $E_{t,T}^{\ast}\leq0$ $\forall t$. As $\alpha>0$, $1-2 \alpha E_{t,T}^{\ast}>0$ $\forall t$.
	
	\hfill $\square$
\end{pf}

As an alternative to (\ref{OptAllocation_periodt 1}) the formula for the
optimal solution can also be decomposed into two different terms in the
following way:
\begin{align}
\pi_{t}^{\ast} = \frac{\lambda +\frac{1}{2}}{1-\gamma}+\frac{\left[%
	(1-\gamma)\left(\theta+(\lambda +\frac{1}{2})\right)+(\lambda +\frac{1}{2})%
	\right]2\alpha E_{t+1,T}^{\ast}}{(1-\gamma)\left(\gamma -(1-2\alpha
	E_{t+1,T}^{\ast})\right)}.  \label{opt pi Merton plus 1}
\end{align}
One of these terms is constant, proportional to the risk
premium and inversely proportional to the investor's risk preference. This
term is usually called the myopic component of an investment (see for
example \cite{Campbell.1999} or \cite{Kraft*.2005}). The second term changes
over time and is often referred to as the investor's hedging demand against
unfavorable relative price changes in assets. This interpretation was first
proposed by \cite{Merton.1973}. In other words, the myopic investor only
makes single-period decisions disregarding future reinvestment opportunities
(see for example \cite{Quek.2017}). As our optimal decision $\pi_{t}^{\ast}$
depends on the time to maturity our strategy is clearly non-myopic.

Our time-dependent hedging component vanishes as the investment horizon $T-t$
goes to $1$. The solution for this case is the solution for the extreme case
where the number of trading periods in between $t$ and $T$ goes to one i.e.
the opposite case to the continuous-time limit where the number of trading
periods goes to infinity. This solution is derived explicitly in the first
part of the proof of Theorem \ref{myTheorem 1} in \ref{proofSpec 1} and
coincides with the myopic term of the above formula. Comparing this to
Merton's solution from \cite{RobertC.Merton.1996}, that is the risk premium
of the stock divided by one minus the risk-aversion parameter, we see that
both are exactly the same. Thus our solution incorporates Merton's as a
particular case. Moreover, there is a second case in which the hedging term
is zero and where we recover Merton's myopic solution. This occurs when $%
\alpha = 0$, which means that the variance is deterministic.

Of paramount importance to an investor is the evolution of his wealth over
time. Applying the results from Theorem \ref{myTheorem 1} to the optimal log
wealth process $\left\{W_{t}^{\ast}\right\}_{t=0}^{T}$ reads
\begin{align}
W_{t}^{\ast}&=W_{t-1}^{\ast} + \left((\lambda +\frac{1}{2})\pi _{t-1}^{\ast}-%
\frac{1}{2}(\pi _{t-1}^{\ast})^{2}\right)h_{t} + \pi _{t-1}^{\ast}\sqrt{h_{t}%
}z_{t}+r  \label{W_rec_opt 1} \\
h_{t} &=\omega +\beta h_{t-1}+\alpha (z_{t-1}-\theta \sqrt{h_{t-1}})^{2}.
\notag
\end{align}
This process has some interesting properties that can be shown analytically.
First of all, we note that the log-wealth process is an affine GARCH process
itself. Its properties are similar to the HN-GARCH model. Effects like
volatility clustering or skewness and kurtosis of returns over multiple
periods can be captured. Corollary \ref{myCorollary 1} presents the moment
generating function of the optimal log-wealth process. With this, all
moments of the distribution can be calculated at any given time $t$.

\begin{cor}
\label{myCorollary 1}  The optimal log-wealth process from (\ref{W_rec_opt 1}%
) is an affine GARCH model. Its conditional moment generating function $%
\mathbb{E}_{t}\left[e^{uW_{T}^{\ast}}\right]$ is given by
\begin{align}
\Psi _{W_{T}^{\ast}}^{\left( t\right)}\left( u\right) &= \mathbb{E}_{t}[e^{u
W_{T}^{\ast}}]=\exp \left\{ uW_{t}^{\ast}+A_{t,T}+B_{t,T}h_{t+1}\right\}
\label{mgfW 1}
\end{align}
where
\begin{align}
A_{t,T} &= A_{t+1,T}+ur+B_{t+1,T}\omega -\dfrac{1}{2}\log (1-2\alpha
B_{t+1,T}),\quad t=0,\dots ,T-1 \\
B_{t,T} &=u\left((\lambda +\frac{1}{2})\pi _{t}^{\ast}-\frac{1}{2}(\pi
_{t}^{\ast})^{2}\right) +\left(\beta+\alpha \theta ^{2}\right) B_{t+1,T}+%
\dfrac{(u\pi_{t}^{\ast}-2\alpha \theta B_{t+1,T})^{2}}{2(1-2\alpha
B_{t+1,T}\,)},\quad t=0,\dots,T-1
\end{align}
with $A_{T,T}=0$, $B_{T,T}=0$.
\end{cor}

\begin{pf}
	See the complimentary material. %See \ref{proof mgfW 1}.
	
	\hfill $\square$
\end{pf}
The log-wealth process is of similar type as the HN-GARCH model but with drift and
variance depending on $\pi_{t}^{\ast}$. Further, this m.g.f. incorporates
the HN-GARCH model as a special case, i.e. setting $\pi_{t}^{\ast} = 1$ for
all $t$. Furthermore, the same stationarity condition for the variance
process applies as in the HN-GARCH model. The return process of the log
wealth though, might not be stationary due to it's dependence on $%
\pi_{t}^{\ast}$.

As an alternative to using the m.g.f. one can also calculate the
multi-period expectation of the optimal wealth process via the following
corollary.

\begin{cor}\label{cor:multiperiodExp}
The multi-period expectation of the optimal log-wealth process (\ref%
{W_rec_opt 1}) is given by
\begin{align*}
&\mathbb{E}_{0}[W_{t}^{\ast}] = w_{0}+rt\\
&+ \sum_{k=1}^{t} \left[\left( (\lambda +\frac{1}{2})\pi _{t-k}^{\ast}-%
\frac{1}{2}(\pi _{t-k}^{\ast})^{2}\right)\left( \left(\alpha + \omega
\right) \left(\frac{1-\left(\beta +\alpha \theta ^{2}\right)^{(t+1-k)}}{%
1-\left(\beta +\alpha\theta ^{2}\right)}\right) + (\beta + \alpha
\theta^{2})^{(t+1-k)}h_{0}\right)\right].
\end{align*}
\end{cor}

\begin{pf}
See the complimentary material.

	\hfill $\square$
\end{pf}

\subsection{Continuous-time limit of the optimal strategy \label{sec:cont
time 1}}

As shown in \cite{Badescu.2019}, the continuous-time limit of the HN-GARCH
model is the model from \cite{Heston.1993} with $\rho = -1$.
Therefore, one would intuitively expect that the limit of the solution under
the HN-GARCH model coincides with the solution under the Heston model. The optimization
problem however, does not only consist of the dynamics of the risky asset.
In order to show that our solution converges to the Heston solution we need
to show that the problems are equivalent as a whole. In general both
problems consist of two components: (i) the utility function and (ii) the
stochastics of $V_{T}$.

The utility function in both problems is the same and given by $U(v) = \frac{%
v^{\gamma}}{\gamma}$. So showing that our solution converges to the Heston
solution from \cite{Kraft*.2005} boils down to showing that the
discrete-time wealth process $V_{t}=e^{W_{t}}$ with
\begin{align}
W_{t}-W_{t-\Delta} &=r(\Delta) + \pi _{t-\Delta}\lambda(\Delta )
h_{t}+\left( \pi _{t-\Delta}-\pi _{t-\Delta}^{2}\right) \frac{1}{2}h_{t}+\pi
_{t-\Delta}\sqrt{h_{t}}z_{t}  \label{W delta} \\
h_{t+\Delta} &= \omega_{h}(\Delta ) +\beta_{h}(\Delta )
h_{t}+\alpha_{h}(\Delta ) (z_{t}-\theta_{h}(\Delta ) \sqrt{h_{t}})^{2},
\label{h delta}
\end{align}
where $\Delta$ represents the time increments, converges to the
continuous-time process
\begin{align}
\frac{dV_{t}}{V_{t}} &= \left( r+\pi_{t}\overline{\lambda }v_{t}\right)
dt+\pi_{t}\sqrt{v_{t}}dz^{S}_{t} \\
dv_{t} &= \kappa \left( \theta _{V}-v_{t}\right) dt-\sigma \sqrt{v_{t}}%
dz^{S}_{t}  \label{dv}
\end{align}
where $z^{S}_{t}$ denotes a standard Wiener process, for $\Delta \rightarrow
0$ i.e. to the portfolio process under the Heston model with $\rho=-1$. This
approach of showing the converges of discrete-time solutions as $\Delta
\longrightarrow 0$ was used earlier by \cite{Rodkina.2016}. They show that
under mild assumptions Merton's continuous-time solution is asymptotically
optimal under a discrete-time model if this model converges to Merton's GBM
as the time steps become very small.

Analogue to \cite{EscobarAnel.2020} we derive the continuous-time limit by
using the weak convergence of Markov processes to diffusions. For this we
assume that $\underset{\Delta \rightarrow0}{\lim}\frac{r(\Delta)}{\Delta}=r$
where $r$ is the instantaneous short rate and start by writing (\ref{W delta}%
) and (\ref{h delta}) in terms of $v_{t} = \frac{h_{t}}{\Delta}$:
\begin{align}
W_{t}-W_{t-\Delta} &=r(\Delta) + \pi _{t-\Delta}\lambda(\Delta )
v_{t}\Delta+\left( \pi _{t-\Delta}-\pi _{t-\Delta}^{2}\right) \frac{1}{2}%
v_{t}\Delta+\pi _{t-\Delta}\sqrt{v_{t}}\sqrt{\Delta}z_{t}  \label{W delta v}
\\
v_{t+\Delta} &= \omega(\Delta ) +\beta(\Delta ) v_{t}+\alpha(\Delta )
(z_{t}-\theta(\Delta ) \sqrt{v_{t}})^{2},  \label{v delta}
\end{align}
with $\omega(\Delta)\coloneqq\frac{\omega_h(\Delta)}{\Delta}$, $\beta(\Delta)\coloneqq\beta_h(\Delta)$, $\alpha(\Delta)\coloneqq\frac{\alpha_h(\Delta)}{\Delta}$, $\theta(\Delta)\coloneqq \theta_h(\Delta)\sqrt{\Delta}$.

The convergence of (\ref{v delta}) to (\ref{dv}) is already known from \cite%
{EscobarAnel.2020}.

To show the convergence of the wealth process itself, it is convenient to
rewrite the continuous-time process in the following way. First, using It%
\^{o}'s Lemma we write the process in terms of log-prices which reads
\begin{align}
dW_{t} &= \left( r+\pi_{t}\overline{\lambda } v_{t}-\frac{1}{2}%
\pi_{t}^{2}v_{t}\right) dt+\pi_{t}\sqrt{v_{t}}dz^{S}_{t}  \notag \\
&= \left( r+\pi_{t}\lambda v_{t}+(\pi_{t} -\pi_{t}^{2})\frac{1}{2}%
v_{t}\right) dt+\pi_{t}\sqrt{v_{t}}dz^{S}_{t}.  \label{W Heston}
\end{align}
where $\bar{\lambda}$, the risk premium on the return, relates to the risk
premium on the log-return $\lambda$ via $\bar{\lambda} = \lambda+\frac{1}{2}$%
. With that one can derive the following proposition.
\begin{prop}\label{myProp}
	The stochastic difference equation (\ref{W delta v}) converges weakly to the stochastic differential equation (\ref{W Heston}).
\end{prop}	
\begin{pf}
	See the complimentary material. %See \ref{proof mgfW 1}.
	
	\hfill $\square$
\end{pf}

This shows that our solution converges to Heston's for $\Delta
\longrightarrow0$. Further our solution is a generalization of Heston's
solution with respect to $\Delta$. From the perspective of a practitioner
that can not rebalance continuously but at discrete points in time, this
additional flexibility of the model is valuable because he can optimize his
strategy exactly to his rebalancing behavior. In a continuous-time model on
the other hand, the investor would suffer losses by implementing the optimal
strategy in a suboptimal way i.e. by rebalancing only discretely. We will
study this point in the numerical section.
%This loss increases with increasing length of the rebalancing periods. Note, that the approximation error of our SFC also rises for larger time intervals. However, this is a matter of accuracy and not a matter of suboptimality as the difference between the optimal wealth process and the actual wealth process can be positive or negative as we will show in Section \ref{sec:proxy 1}.

An alternative treatment of the convergence of discrete-time solutions using
ordinary differential equations can be found in \cite{Bensoussan.2014}.

\subsection{Losses from suboptimal strategies \label{sec:subopt}}

In this subsection we study the wealth-equivalent loss (WEL) an investor
suffers by following a suboptimal strategy $\pi^{s}$. Furthermore we present
some properties of suboptimal strategies. The expected utility from terminal
wealth for an investor following such a strategy can be written analogously
to Section \ref{sec:Setting general} while omitting the maximization, i.e.
only using the tower property of conditional expectations. This is,
\begin{align}
\Phi_{0}^{s} (\log (v_{0}),h_1) &= \mathbb{E}_{0}\left[ U\left(\log (V_{T})\right) \right] =%
\mathbb{E}_{0}\left[\mathbb{E}_{1}\left[ \dots\mathbb{E}_{T-2}\left[ \mathbb{E}%
_{T-1}\left[ U\left(\log (V_{T})\right) \right] \right] \dots \right] \right]
\end{align}
and
\begin{align}
\Phi^{s}_{t}(\log(V_{t}),h_{t+1}) &= \mathbb{E}_{t}[\Phi^{s}_{t+1}(\mathbb{T}(\log(V_t),\pi_{t}^{s},h_{t+1},Y_{t+1}))],
\end{align}
with $\Phi^{s}_{T}(\log(V_{T}),h_{T+1}) = \Phi^s_T(\log(V_T))= U(\log(V_{T}))$.\\
From the definition of $\Phi_t(V_{t},h_{t+1})$ in (\ref{EUT_rec}) it follows that $%
\Phi_t^{s}(V_{t},h_{t+1}) \leq \Phi_t(V_{t},h_{t+1})$ with equality when $\pi^{s}=\pi^{\ast}$.

Following \cite{Escobar.2015} we define the wealth-equivalent utility loss $%
L_{t}^{s}$ from following a suboptimal strategy as the solution to
\begin{align}
\Phi_t(\log(V_{t}(1-L_{t}^{s})),h_{t+1}) &= \Phi_t^{s}(\log(V_{t}),h_{t+1}).  \label{WEL_impl}
\end{align}

An investor following the optimal strategy thus only needs a fraction of $1-L_{t}^{s}$ of the initial capital to achieve the same expected utility as if he applies the suboptimal strategy. In other words, applying the suboptimal strategy, a fraction of $L_{t}^{s}$ of the initial capital would be wasted or lost compared to the optimal strategy.

To arrive at an explicit expression for $L_{t}^{s}$ some preparation is
needed. Let us denote the set of admissible strategies $\mathcal{U}[0,T]$ additionally including the
condition $1-2 \alpha E_{t,T}^{s}>0 \quad \forall t\in[0,T]$, with $E_{t,T}^{s}$ as per equation (\ref{E_pert sub}), by $\mathcal{U}^{s}[0,T]$.\footnote{The last condition is a technical one assuring that the formulas obtained in
this section are well defined.}

\begin{prop}
\label{myProp1}  For any admissible strategy $\pi_{t}^{s}$ the expected
utility from terminal wealth conditioned on $t$ is given by
\begin{align}
\Phi_t^{s} (W_{t},h_{t+1})&=\frac{1}{\gamma }\exp \left\{ D_{t,T}^{s}+\gamma
W_{t}+E_{t,T}^{s}h_{t+1}\right\}  \label{Value_pert 1}
\end{align}
where
\begin{align}
D_{t,T}^{s} &= D_{t+1,T}^{s} + E_{t+1,T}^{s}\omega+\gamma r-\log \left(
\sqrt{(1-2\alpha E_{t+1,T}^{s})}\right),\quad t=0,\dots,T-1  \label{A_pert1}
\\
E_{t,T}^{s} &= \left( \beta +\alpha \theta ^{2}\right)E_{t+1,T}^{s} +\dfrac{%
\left( \gamma \pi_{t}^{s}-2\theta \alpha E_{t+1,T}^{s}\right) ^{2}}{%
2(1-2\alpha E_{t+1,T}^{s})}+ \gamma \left((\lambda + \frac{1}{2})\pi_{t}^{s}-%
\frac{1}{2}(\pi_{t}^{s})^{2}\right),\quad t=0,\dots,T-1  \label{E_pert sub}
\end{align}
with $D_{T,T}^{s} = E_{T,T}^{s} = 0$.
\end{prop}

\begin{pf}
See the complimentary material. 	
%This follows by applying the tower rule of expectations which is equivalent to following along the lines of the proofs in \textit{\ref{proofSpec 1}} while skipping the optimization step.
	
	\hfill $\square$
\end{pf}

Now we can derive an explicit expression for $L_{t}^{s}$ which is presented
in the following lemma.

\begin{lem}
\label{lem WEL}
\begin{align*}
L_{t}^{s}(h_{t+1}) &= 1 - \exp\left\{\frac{1}{\gamma}\left(\left(D_{t,T}^{s}-D_{t,T}%
\right) + \left(E_{t,T}^{s}-E_{t,T}^{\ast }\right)h_{t+1}\right)\right\}
\end{align*}
\end{lem}

\begin{pf}
See the complimentary material.
	
	\hfill $\square$
\end{pf}

These results will be used later to conduct numerical experiments that
compare meaningful suboptimal and optimal strategies with respect to WEL.

Further note that the m.g.f. of the log-wealth process can be found for any
admissible strategy $\left\{\pi_{t}^{s}\right\}_{t\in\left\{0,1,\dots
T\right\}} \in \mathcal{U}^{s}[0,T]$ as this is an affine GARCH model as well. This is shown
in Proposition \ref{myProp2} and is a useful result. Imagine an investor
decides to construct a strategy that is different from the optimal one in
this paper. The strategy is of course suboptimal in our setting but the
investor might want to study properties of such a suboptimal portfolio for
risk management or pricing purposes. In this case the m.g.f. provides a
quick way to calculate risk measures and moments for any strategy he
considers without the need for lengthy simulations. Furthermore, as this
GARCH is of the Heston-Nandi type one can find a risk-neutral representation
of the model. This makes it possible to price derivatives on such a
portfolio.

\begin{prop}
\label{myProp2}  For any admissible strategy $\pi_{t}^{s}$ the log-wealth
process is an affine GARCH model. Its conditional moment generating function
$\mathbb{E}_{t}\left[e^{uW_{T}}\right]$ is given by
\begin{align}
\Psi _{W_{T}}^{\left( t\right)}\left( u\right) &= \mathbb{E}_{t}[e^{u
W_{T}}]=\exp \left\{ uW_{t}+A_{t,T}+B_{t,T}h_{t+1}\right\}  \label{mgfW sub}
\end{align}
where
\begin{align}
A_{t,T} &= A_{t+1,T}+ur+B_{t+1,T}\omega -\dfrac{1}{2}\log (1-2\alpha
B_{t+1,T}),\quad t=0,\dots ,T-1 \\
B_{t,T} &=u\left((\lambda +\frac{1}{2})\pi _{t}^{s}-\frac{1}{2}(\pi
_{t}^{s})^{2}\right) +\left(\beta+\alpha \theta ^{2}\right) B_{t+1,T}+\dfrac{%
(u\pi_{t}^{s}-2\alpha \theta B_{t+1,T})^{2}}{2(1-2\alpha B_{t+1,T}\,)},\quad
t=0,\dots,T-1
\end{align}
with $A_{T,T}=0, B_{T,T}=0$.
\end{prop}

\begin{pf}
See the complimentary material.
	%The proof follows along the lines of the one in \ref{proof mgfW 1} while plugging in any general $W_{t}$ instead of $W_{t}^{\ast}$.
	
	\hfill $\square$
\end{pf}

\section{Numerical analysis\label{sec:second order num}}

In this section we present some numerical results on the optimization
approach presented above. Subsection \ref{sec:proxy 1} analyzes the impact
of the approximation used in the derivation of the SFC while Section \ref%
{sec:solSens} discusses the sensitivity of the optimal solution to various
parameters. Section \ref{sec:SolComp} compares our optimal allocation to
other well-known solutions of the portfolio optimization problem. Finally,
in Section \ref{sec:solPerf} we compare the performance of our strategy to
other strategies in terms of wealth-equivalent losses.

Throughout this section we consider the parameter estimates of daily returns
from \cite{Christoffersen.2006} for the HN-GARCH model:
\begin{equation*}
 \alpha = 3.660\cdot 10^{-6}, \quad \beta = 0.9026, \quad \lambda = 2.772, \quad
\theta = 128.4, \quad \omega = 3.038\cdot 10^{-9}
\end{equation*}

\subsection{On the approximation of the SFC}
\label{sec:proxy 1}

In Subsection \ref{sec:Setting general} we derived (\ref{SFC}) as the exact
SFC for the wealth process. Working with log prices, $X_{t}=\log S_{t}$ and
using that $r_{d} = e^{r}-1$ where $r_{d}\coloneqq \frac{B_{t}-B_{t-1}}{%
B_{t-1}}$ is the discrete return and $r\coloneqq \log \frac{B_{t}}{B_{t-1}}$
the log-return of the risk-free bond, we arrive at%
\begin{align}
V_{t} &= V_{t-1}\left(\pi _{t-1}e^{X_{t}-X_{t-1}}+\left( 1-\pi _{t-1}\right)
e^{r} \right).  \label{exact Dynamics 1}
\end{align}
As a proxy for this process we worked with
\begin{align}
W_{t} &= W_{t-1}+\pi _{t-1}\left( X_{t}-X_{t-1}\right) +\left( 1-\pi
_{t-1}\right) r +\left( \pi _{t-1}-\pi _{t-1}^{2}\right) \frac{1}{2}h_{t}
\notag \\
\Leftrightarrow V_{t} &= V_{t-1} \exp \left\{\pi _{t-1}\left(
X_{t}-X_{t-1}\right) +\left( 1-\pi _{t-1}\right) r+\left( \pi _{t-1}-\pi
_{t-1}^{2}\right) \frac{1}{2}h_{t}\right\}.  \label{proxy Dynamics 1}
\end{align}
While the optimal wealth process that is generated by the strategy proposed
in this paper follows the dynamics in equation (\ref{proxy Dynamics 1}) the
self-financing dynamics of the portfolio are given by equation (\ref{exact
Dynamics 1}). To assess this difference we perform simulations with the
following configuration:
\begin{align*}
r &= 0.01/252, \quad \gamma = -5, \quad T = 5\cdot252 \text{ days (i.e. 5 trading years)}, \quad \text{%
number of simulations: } 10,000
\end{align*}

\begin{figure}[t]
\caption{Distribution of terminal wealth for the proxy and exact wealth
process}
\label{fig:TrueProxyDiffVT}
\begin{center}
\includegraphics[width=0.6\linewidth,
height=.3\textheight]{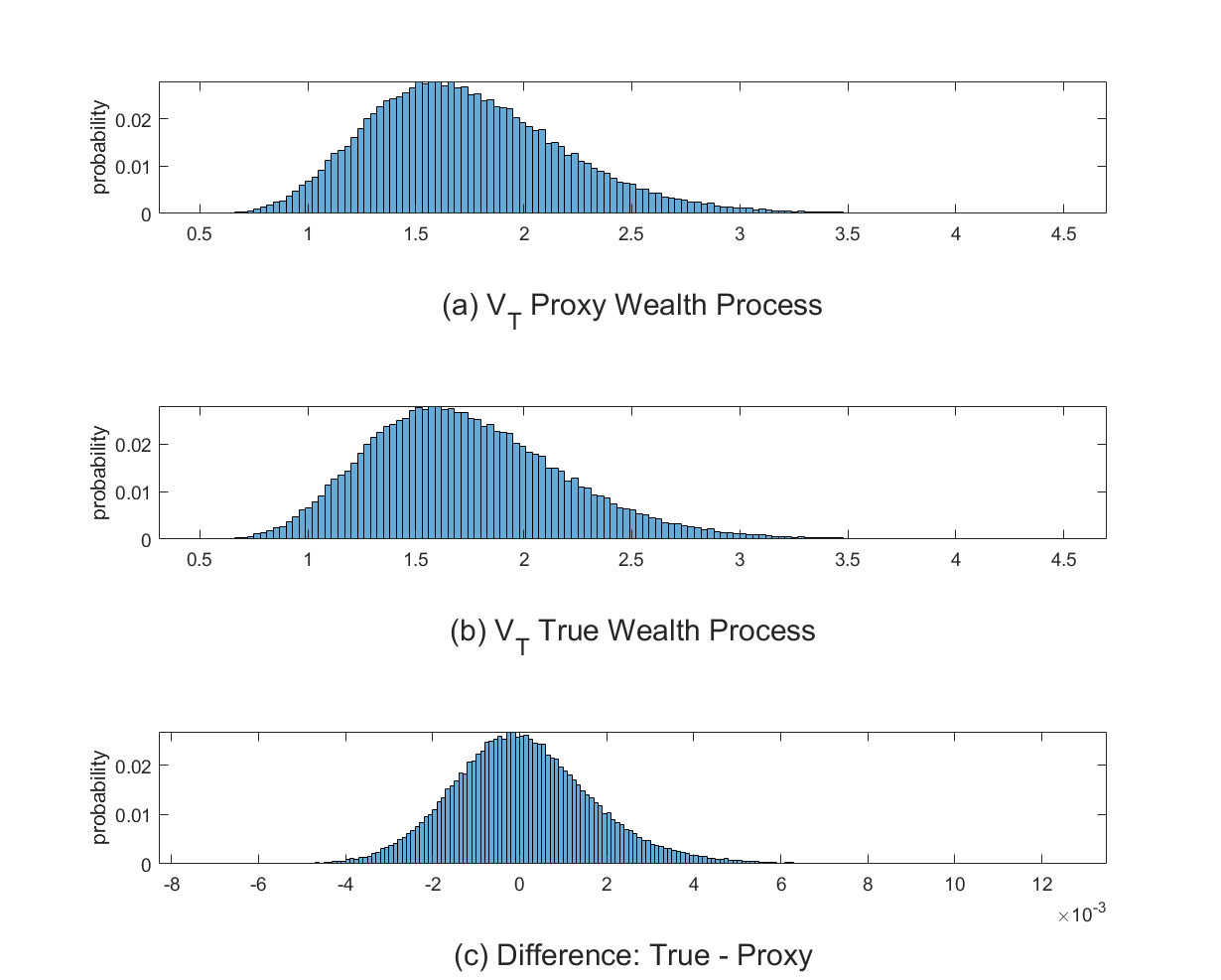}
\end{center}
\par
\noindent {\footnotesize This figure plots the distribution of terminal
wealth after a 5-year investment horizon for the approximated wealth process
(a) and the exact wealth process (b). Further, it plots the distribution of
the difference in terminal wealth after the same horizon in subplot (c).}
\end{figure}

An important question for someone who wants to implement the proposed
strategy is whether the performance of the portfolio is compromised when
using (\ref{proxy Dynamics 1}) as an approximation for (\ref{exact Dynamics
1})? To answer this question we simulated pairs of sample paths of the
wealth process generated by the optimal strategy. The two paths of one pair
were calculated by means of (\ref{exact Dynamics 1}) (a) and (\ref{proxy
Dynamics 1}) (b) respectively. Both use the same realization of the random
variable. In Figure \ref{fig:TrueProxyDiffVT} we report the distribution of
the terminal wealth for both processes (subplots (a) and (b)). Looking at
these two subplots we see that both distributions are almost identical.
Subplot (c) presents the distribution of the difference in terminal wealths.
The center of the distribution is close to $0$. The difference in terminal
wealth one might encounter after a 5-year investment period is mostly in the
range from $-0.4\%$ to $0.6\%$ of the initial investment.

\begin{figure}[t]
\caption{Distribution of cash supplied to maintain the optimal wealth process
}
\label{fig:CashHist}
\begin{center}
\subfigure[Daily cash
flows]{\includegraphics[width=\size\linewidth]{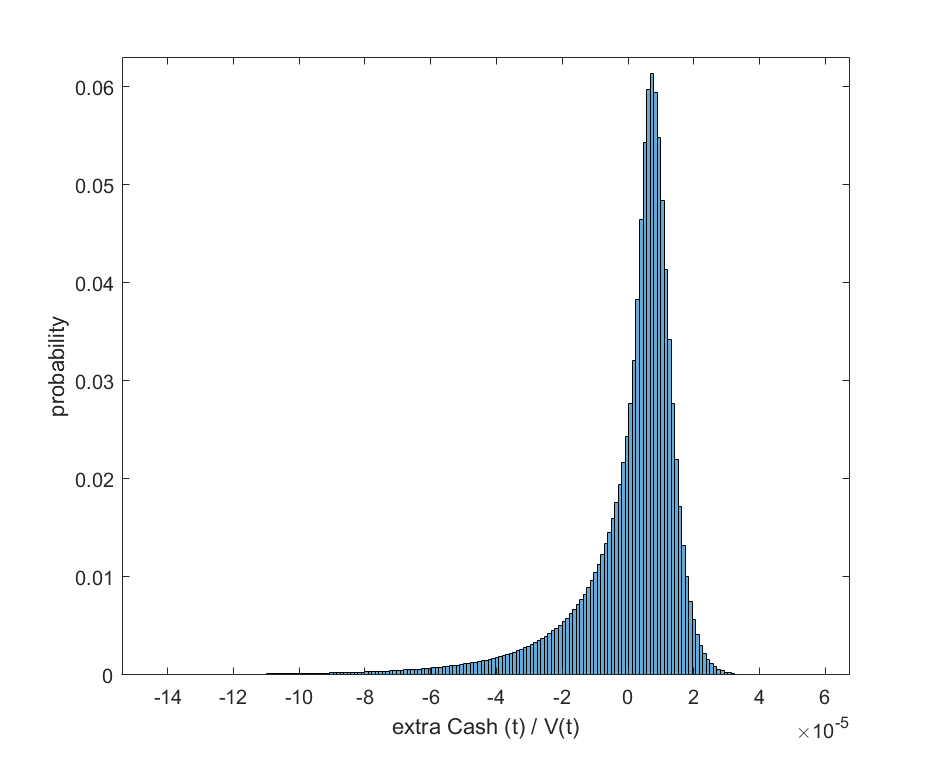}}  %
\subfigure[5-Year cash
flows]{\includegraphics[width=\size\linewidth]{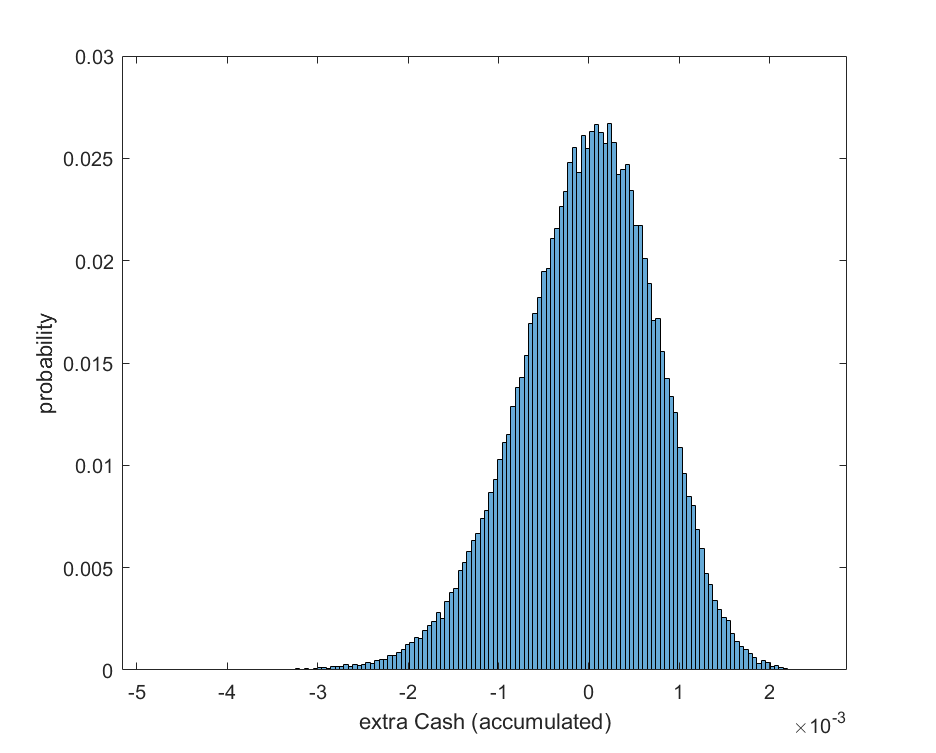}}
\\[0pt]
\end{center}
\par
\noindent {\footnotesize This figure plots the distribution of the extra
cash flows that are necessary to compensate the fact that the approximated
wealth process in not exactly self-financing. Subplot (a) displays the
distribution of daily cash flows as a fraction of the current wealth.
Subplot (b) displays the distribution of the cumulative daily cash flows
over a 5-year horizon.}
\end{figure}

If an investor wants to follow the optimal wealth process exactly he needs
to either pay cash into the strategy or withdraw cash from the strategy in
every period i.e. in every period he needs to set $V_{t-1} = \exp(W_{t-1})$.
The distribution of the cash flows necessary to maintain such a
non-self-financing strategy is shown in Figure \ref{fig:CashHist}. The
number of simulations for this figure was $100,000$. Subplot (a) displays
the distribution of daily cash flows. We observe that the distribution is
negatively skewed. This means that in most cases the investor has to supply
a small amount of money and in fewer cases he can withdraw a relatively high
amount. Most cash flows that need to be supplied to maintain the optimal
strategy are in a range from $-0.008\%$ to $0.003\%$ of the investors
current wealth. Subplot (b) presents the distribution of the accumulated
daily cash flows over the investment horizon of 5 years. It is negatively
skewed as well but less so than the distribution in subplot (a). The
magnitude of the cumulative cash flows supplied over 5 years is of course
somewhat larger than the one of the daily cash flows and lies in a range
between $-0.3\%$ and $0.25\%$ of the investors initial wealth. As we will
discuss the optimality of our solution for $\gamma \in (0,1]$ later we also
performed all calculations for this case. The absolute magnitude of the
cash flows was very similar to what is shown here. The distribution seems to
be the same but mirrored on a vertical axis at zero i.e. on average cash has
to be supplied to the strategy.

Overall, we conclude that the approximation of the SFC that was used to
derive the optimal strategy has only a minor influence on the evolution of
the investors wealth.

\subsection{Sensitivity of the optimal solution \label{sec:solSens}}

\begin{figure}[t]
\caption{Sensitivity of the optimal solution to various parameters}
\begin{center}
\includegraphics[width=.9\linewidth,
height=0.6\textheight]{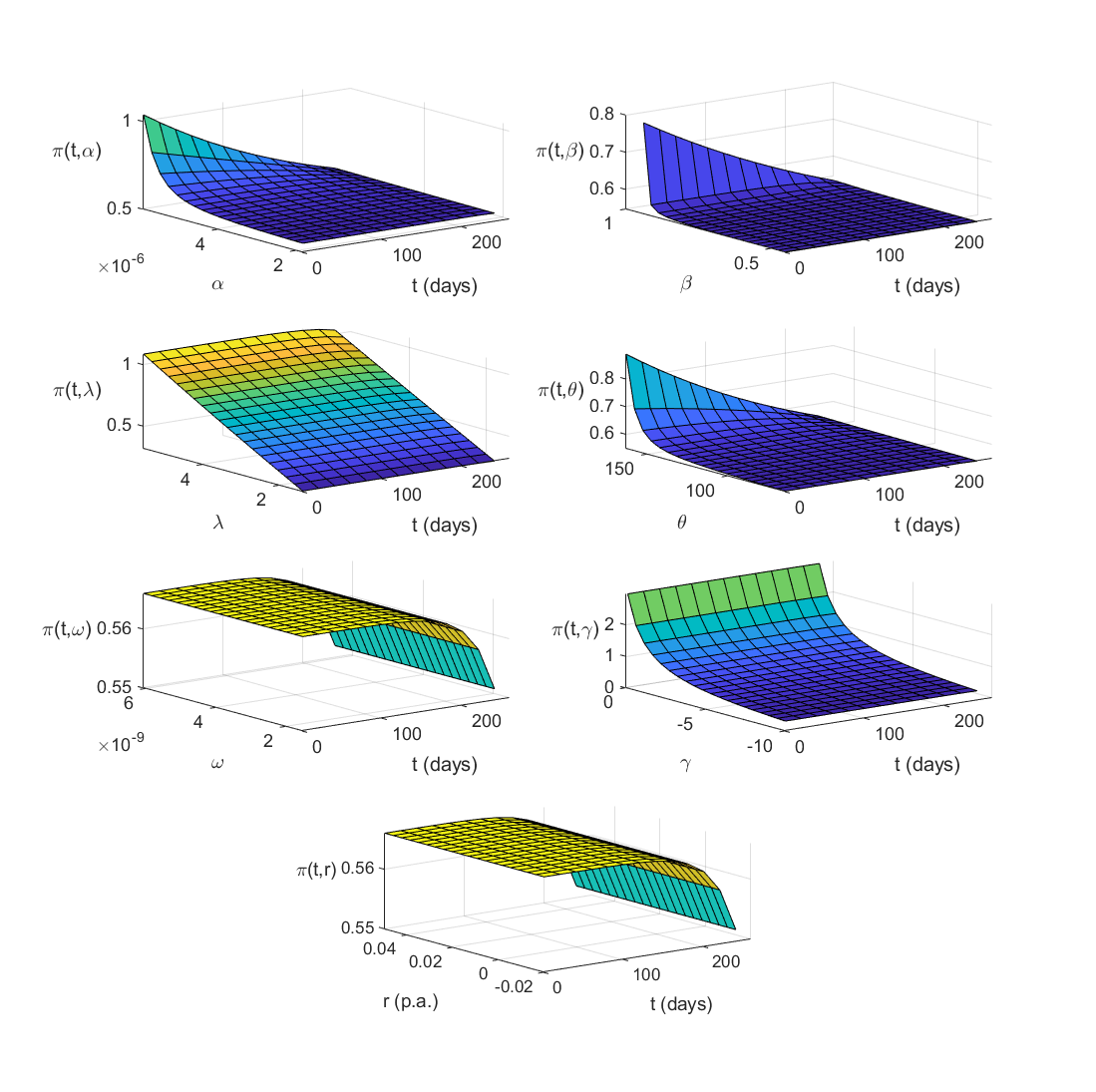}
\end{center}
\par
\noindent {\footnotesize This figure plots the sensitivity of the optimal
investment strategy to changes of various parameters.}
\label{fig:sensParam}
\end{figure}

In this subsection we investigate how the optimal strategy reacts to changes
in model parameters. The optimal solution was calculated for $T=252$ trading
days and model parameters in a range from 50\% to 200\% of the estimates in
\cite{Christoffersen.2006}. While varying one parameter all others were held
constant to 100\% of the \cite{Christoffersen.2006} values with $\gamma=-5$.
The parameters $\alpha$, $\beta$ and $\theta$ where capped such that the
stationarity condition of the HN-GARCH model i.e. $\beta+\alpha \theta^2<1$
is satisfied at any time. $\gamma$ was in a range from $-0.1$ to $-10$ and $r$
in the range from $-2\%/252$ to $4\%/252$. The results are shown in Figure %
\ref{fig:sensParam}. The first horizontal axis shows the different values of
each parameter, while the second horizontal axis shows the time.

One parameter that stands out is $\lambda$ i.e. the market price of risk.
The fraction invested in the risky asset rises linearly with $\lambda$ for
all $t$. This is what one would expect. An agent who found a utility
maximizing trade-off between risk and reward would not take any more
absolute risk unless the reward per unit of additional risk increases. In
the latter case, the agent would be willing to invest a higher fraction of wealth in the risky asset.
Another obvious picture can be observed for $\gamma$. A lower risk aversion,
which corresponds to a $\gamma$ closer to zero, implies a higher investment
in the risky asset. $\omega$ and $r$ have no significant impact whatsoever.
As for $r$ one can already see from the theoretical results that the optimal
strategy is independent of $r$. $\omega$ is traditionally very small when
estimating GARCH models and has thus only minor impact on the optimal
solution. Often, this parameter is even assumed to be zero.

The more interesting parameters are $\alpha$, $\beta$ and $\theta$. They
have a similar impact on the optimal investment as $\pi^{\ast}$ rises with
parameters increasing to values that just fulfill the stationarity condition
while showing a decreasing sensitivity as $t\longrightarrow T$. $\beta$ is
the parameter that models the intensity of autocorrelation in the variance
process. Thus, lower values of $\beta$ imply stronger mean-reversion of the
variance and vice versa. Furthermore, from (\ref{h}) we know that lower
values of $\beta$ also imply a lower average level of variance. From Figure \ref{fig:sensParam} we can see that it is optimal for an investor to have a higher exposure to the risky asset when $\beta$ is higher i.e. when mean-reversion is weaker and when the average variance is higher. This might seem counter intuitive at first but it becomes clear when digging a little deeper. First, from (\ref{HN log stock}) one can see that if variance increases the expected return increases as well. Thus, for rising values of $\beta$ we have a trade-off between higher variance and higher expected return. Therefore, an increase in $\beta$ does not necessarily imply lower exposure to the risky asset. Second, one has to take correlation into account and note
that changes in $\beta$ only impact the non-myopic hedging term. When
correlation between stock and variance becomes more negative the hedging
effect between the two increases. This is, variance risk and stock risk
cancel out to a certain degree. Hence, overall risk decreases and one can
invest more risky.

This impact of correlation is consistent with the solution under the Heston
model. In the setting of a HN-GARCH model, increasing volatility
implies more negative correlation between stock and variance which can be seen from similar considerations as for equation (\ref{moments1 -1}). Thus, it makes sense that for some
parametric settings an increase in variance increases the non-myopic part of
the optimal allocation. Our results show, that this is indeed the case for
reasonable model parameters. While in the literature $\alpha$ is mainly held
responsible for the kurtosis and $\theta$ for the skewness of multi-period
asset returns, both parameters actually influence both distribution
characteristics. The increase of each of those parameters leads to an
increase in kurtosis and a decrease in skewness i.e. returns become more
negatively skewed. It seems intuitive that with decreasing skewness the
attractiveness of the risky asset declines. For the kurtosis on the other
hand this is not so clear as a higher kurtosis implies a higher probability
of both tails of the distribution. Furthermore, as of equation (\ref{h}) higher values of $\alpha$ or $\theta$ also imply a higher
average level of variance and thus more negative correlation between stock and variance. Hence, similar
arguments hold as for $\beta$. Our results suggest, that the larger right
tail of the returns distribution due to the increase in kurtosis and the
more negative correlation between stock and variance lead to an increase in
the optimal allocation. Hence, $\pi^{\ast}$ rises with increasing values of $%
\theta$ and $\alpha$. Furthermore note, that there might be an
interdependence of the effect of different parameters on the optimal
strategy especially for the parameters $\alpha$, $\beta$ and $\theta$ as
they are linked via the stationarity condition. For example, if $\alpha$ is
smaller then the impact of the same increase in $\beta$ may be smaller as
if $\alpha$ is bigger.

Overall we find that the optimal solution is quite robust against small
changes in parameters. This is, when parameters are miss-specified, for
example due to inaccurate estimation, the resulting strategy is not far off
the optimal one using the true parameter set.

\subsection{Convergence of the optimal strategy to known solutions \label%
{sec:SolComp}}

We already related our solution to the continuous-time ones under the Merton
and Heston models from a theoretical point of view. Now let's enrich this
with some numerical results.

\begin{figure}[t]
\caption{Comparison of different strategies}
\label{fig:stratComp}
\begin{center}
\includegraphics[width=.95\linewidth, height=.35\textheight]{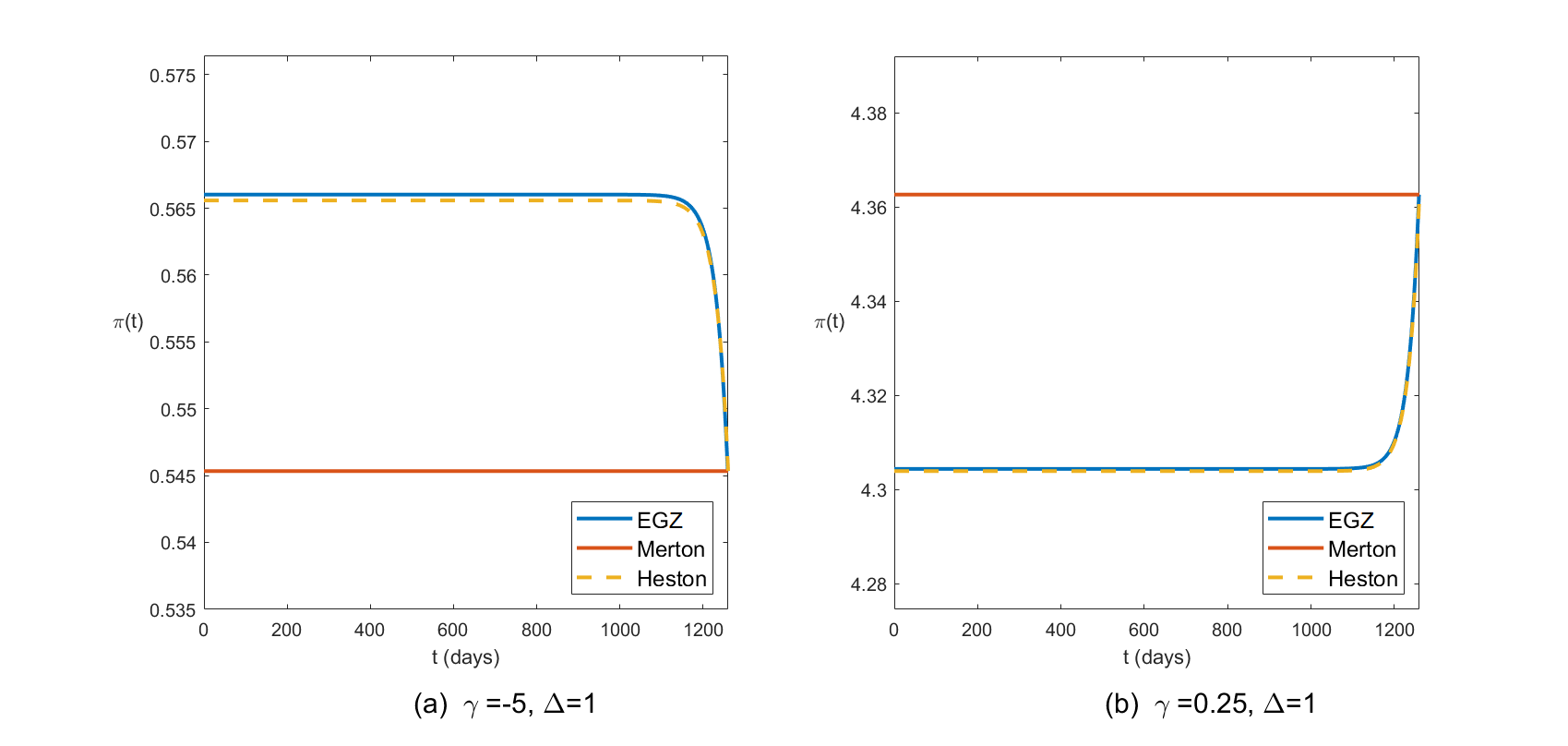}
\end{center}
\par
\noindent {\footnotesize This figure shows a comparison between our strategy
and the Heston and Merton strategy.}
\end{figure}

Figure \ref{fig:stratComp} reports the different portfolio weights over time
for the parameter configuration $r = 0.01/252$, $T = 5\cdot252$ days. In this figure we denote our solution "EGZ". Not
surprisingly, given the daily frequency of the parameters, our solution is
close to the Heston solution regardless of the sign of $\gamma$. Even though
its analytical derivation only shows optimality for $\gamma<0$ explicitly,
this fact supports the hypothesis that our solution is optimal for $\gamma<1$%
, $\gamma\neq0$ similar to both continuous-time solutions.

Both, the Heston solution and our solution are greater than Merton's for $%
\gamma<0$ and smaller than Merton's for $\gamma>0$ over the whole investment
horizon. Overall, all three solutions are rather close to each other for the
selected parameter configurations. Note that if $t \longrightarrow T-1$ our
solution converges to Merton's. This is the same result that we found in
Section \ref{sec:results1}.

\begin{figure}[t]
\caption{Convergence behavior of the optimal solution}
\label{fig:stratCompConvergence}
\begin{center}
\includegraphics[width=.95\linewidth,
height=.5\textheight]{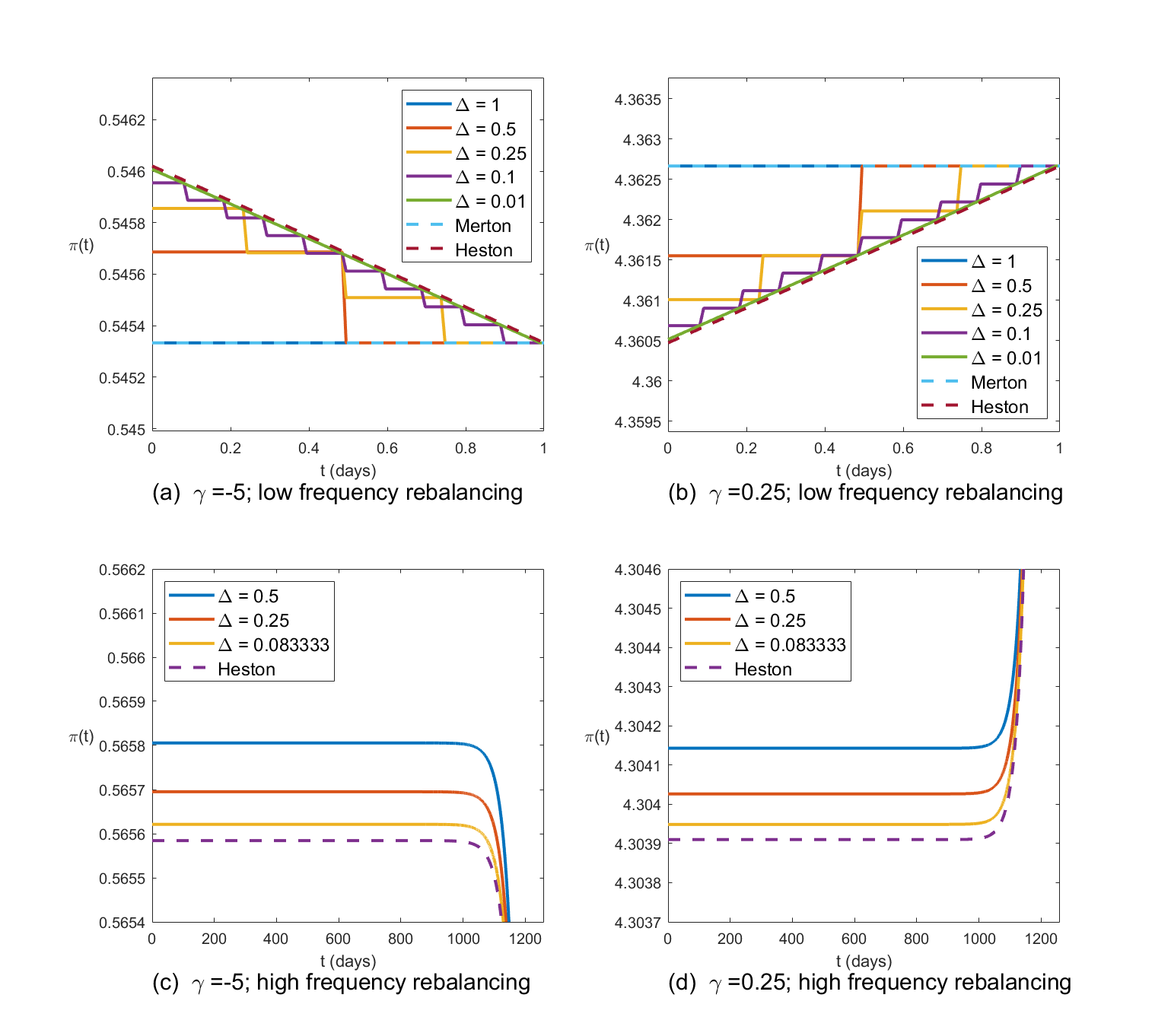}
\end{center}
\par
\noindent {\footnotesize This figure shows convergence behavior of our
optimal solution as the time step goes to zero.}
\end{figure}

Figure \ref{fig:stratCompConvergence} displays the convergence behavior of
our solution as $\Delta\longrightarrow0$. Subplots (a) and (b) show the case
where $T=1$ day. We use this investment horizon to show our solution for low
rebalancing frequencies, i.e. few rebalancings over the fixed horizon. We
have defined the rebalancing frequency as $\frac{T}{\Delta}$ i.e. the number
of rebalancing periods over the investment horizon. From \cite{Badescu.2019}
we know how the parameters change with $\Delta$ for $\Delta\leq1$ day as we
use parameters estimated from daily returns. How the parameters dependent on
$\Delta$ for all $\Delta>1$ day for daily returns is unknown in the
literature and a topic for future research. Note that, for
the purpose of writing our solution dependent on $\Delta$, we need to write
the parameter specifications from \cite{Badescu.2019} in terms of $h_{t}$
instead of $v_{t}$. As $\Delta$ has this upper bound of $1$, we can only
show low rebalancing frequencies $\frac{T}{\Delta}$ by choosing small $T$.
Setting $T=1$, $\Delta=0.5$ implies rebalancing twice, $\Delta=0.25$ implies
choosing weights four times and so on. For $\Delta < T$ our solution departs
from Merton's and moves towards the Heston solution. The behavior is the
same for $\gamma<0$ and $\gamma>0$.

Subplots (c) and (d) in Figure \ref{fig:stratCompConvergence} show the
convergence behavior for high rebalancing frequencies, i.e. many rebalancing
times for the fixed investment horizon. For this we use 5 years i.e. $T=5\cdot252
$ again. In the case $\gamma>0$ (subplot(d)) everything is as suggested by
subplot (b). Our solution converges to the Heston solution for $%
\Delta\longrightarrow0$. Curiously, subplot (c) which shows the case $%
\gamma<0$ suggests a "non-linear" convergence behavior. In most regions of $%
\Delta$ our solution is smaller than Heston's and converges to it from below
for a shrinking $\Delta$. But for very small rebalancing frequencies, our
solution rises above Heston's before converging back to it from above. This
indicates that there are opposite forces at work as $\Delta\longrightarrow0$%
, one that pulls our solution towards the Heston solution and one that
pushes it above. The, latter seems to dominate only for very specific
regions of $\Delta$. An economic explanation for the second force might be
that the discrete-time model assumes that there is no movement of the risky
asset in-between two points in time, while under the continuous-time model
there is movement at any point in time. Thus, the same risky asset might be
perceived as less risky under the discrete-time model and therefore a higher
stake in the risky asset can be chosen for the same risk preference.

\subsection{Performance of the optimal and suboptimal strategies\label%
{sec:solPerf}}

A numerical comparison of the performance for different strategies is
conducted in this subsection. We again use simulations to implement some of
the analysis. The parameters for the simulation are:
\begin{align*}
V_{0} = 1, \quad r = 0.01/252, \quad \gamma = -5, \quad T = 252 \text{ days}%
, \quad \text{number of simulations: } 1,000,000
\end{align*}
Table \ref{tab:moments} reports the moments of the one year return
distribution and the expected utility from terminal wealth generated by
different strategies. The latter was calculated using the closed-form
expressions from Theorem \ref{myTheorem 1}. The same expressions can also be
applied to any other investment strategy as per Lemma \ref{myProp1}.
We denote our optimal strategy "EGZ". "Heston" refers to the strategy from
\cite{Kraft*.2005} while "Merton" refers to the one from \cite%
{RobertC.Merton.1996}. Merton's strategy deviates the most from the other
two. We can see that it produces a slightly higher Sharpe ratio and lower
kurtosis than the other two strategies. On the other hand it also has more
negatively skewed returns and most importantly it obviously has a lower
expected utility from terminal wealth. The Heston solution holds up quite
well in our setting based on this analysis.
\begin{table}[htbp]
\centering
\begin{tabular}{lrrrrrr}
\toprule strategy & \multicolumn{1}{l}{$\mu$} & \multicolumn{1}{l}{$\sigma$}
& \multicolumn{1}{l}{skewness} & \multicolumn{1}{l}{kurtosis} &
\multicolumn{1}{l}{SR} & \multicolumn{1}{l}{$E[U_{T}]$} \\
\midrule EGZ & 0.0567 & 0.0863 & -0.1472 & 3.0546 & 0.6568 & -168.9989E-03
\\
Heston & 0.0566 & 0.0863 & -0.1471 & 3.0471 & 0.6560 & -168.9990E-03 \\
Merton & 0.0552 & 0.0833 & -0.1561 & 3.0494 & 0.6628 & -169.0227E-03 \\
\bottomrule &  &  &  &  &  &
\end{tabular}%
\caption{Distribution moments of realized returns and expected utility from
terminal wealth generated by different strategies. }
\label{tab:moments}
\end{table}

Based on the results from Section \ref{sec:subopt}, we further estimate the
utility loss for an investor who follows a suboptimal strategy. We start by
exploring the WEL induced by following the Heston and the Merton strategy
instead of the strategy optimal to this setting and its sensitivity towards
the risk aversion parameter $\gamma$. The results are shown in Figure \ref%
{fig:WELvsGamma_T10}. Subplot (a) shows the whole region, while subplot (b)
provides a closer look at the comparison to the Heston strategy. The first
thing to note is that, for the parameter set at hand, WEL is relatively
small over all values of $\gamma$. Furthermore, the WEL induced by the
Heston solution is almost negligible compared to the one induced by Merton's
solution.
\begin{figure}[t]
\caption{Wealth-equivalent loss from suboptimal strategies with $T=252*10$
days.}
\label{fig:WELvsGamma_T10}
\begin{center}
\subfigure[] {\includegraphics[width=\size\linewidth]{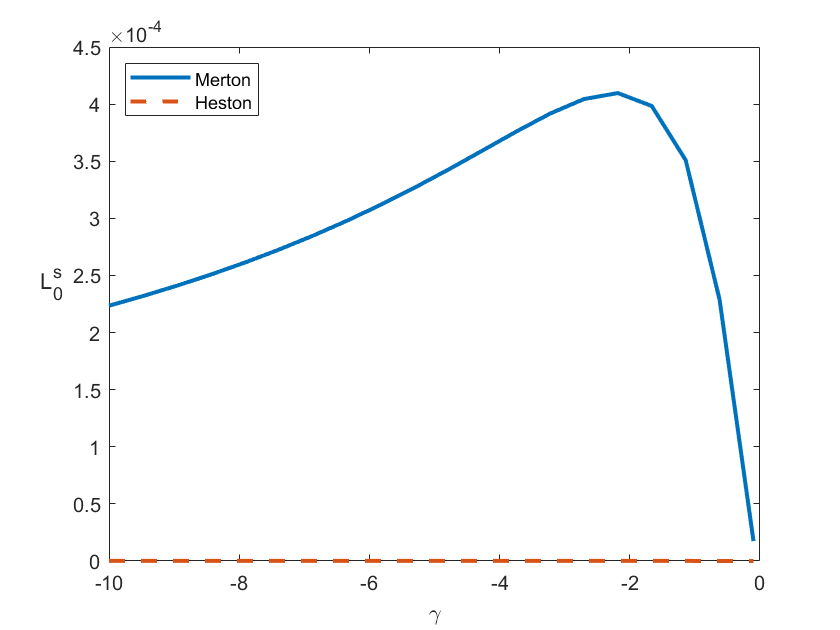}}  %
\subfigure[]
{\includegraphics[width=\size\linewidth]{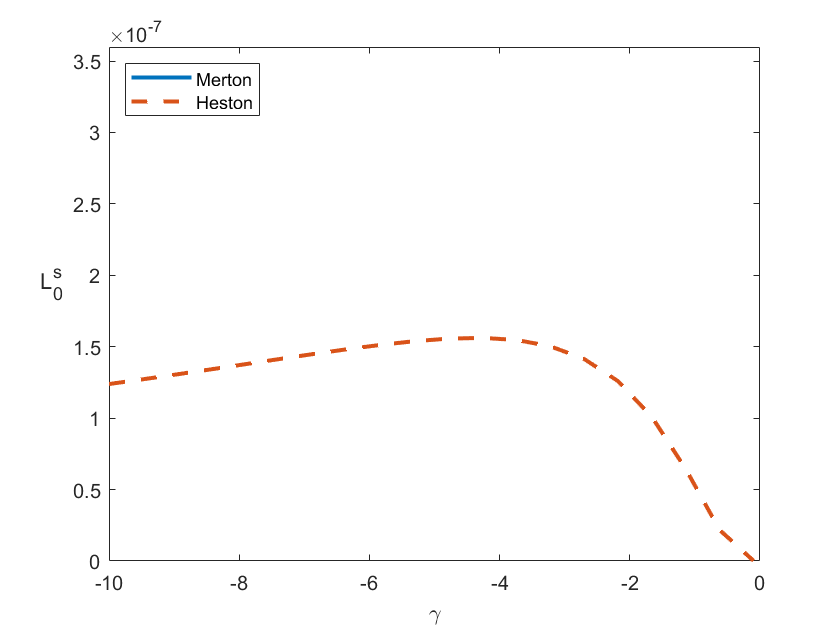}} \\[0pt]
\end{center}
\par
\noindent {\footnotesize This figure plots the WEL induced by following
Merton's and Heston's strategy for different levels of risk aversion $\gamma$%
. Subplot (a) shows the whole picture, subplot (b) provides a closer look at
the comparison to the Heston strategy.}
\end{figure}

\begin{figure}[t]
\caption{Wealth-equivalent loss from following suboptimal strategies with $%
T=252\cdot10$ days and $\protect\gamma = -5$.}
\label{fig:WELvsParams}
\begin{center}
\includegraphics[width=.95\linewidth,
height=.6\textheight]{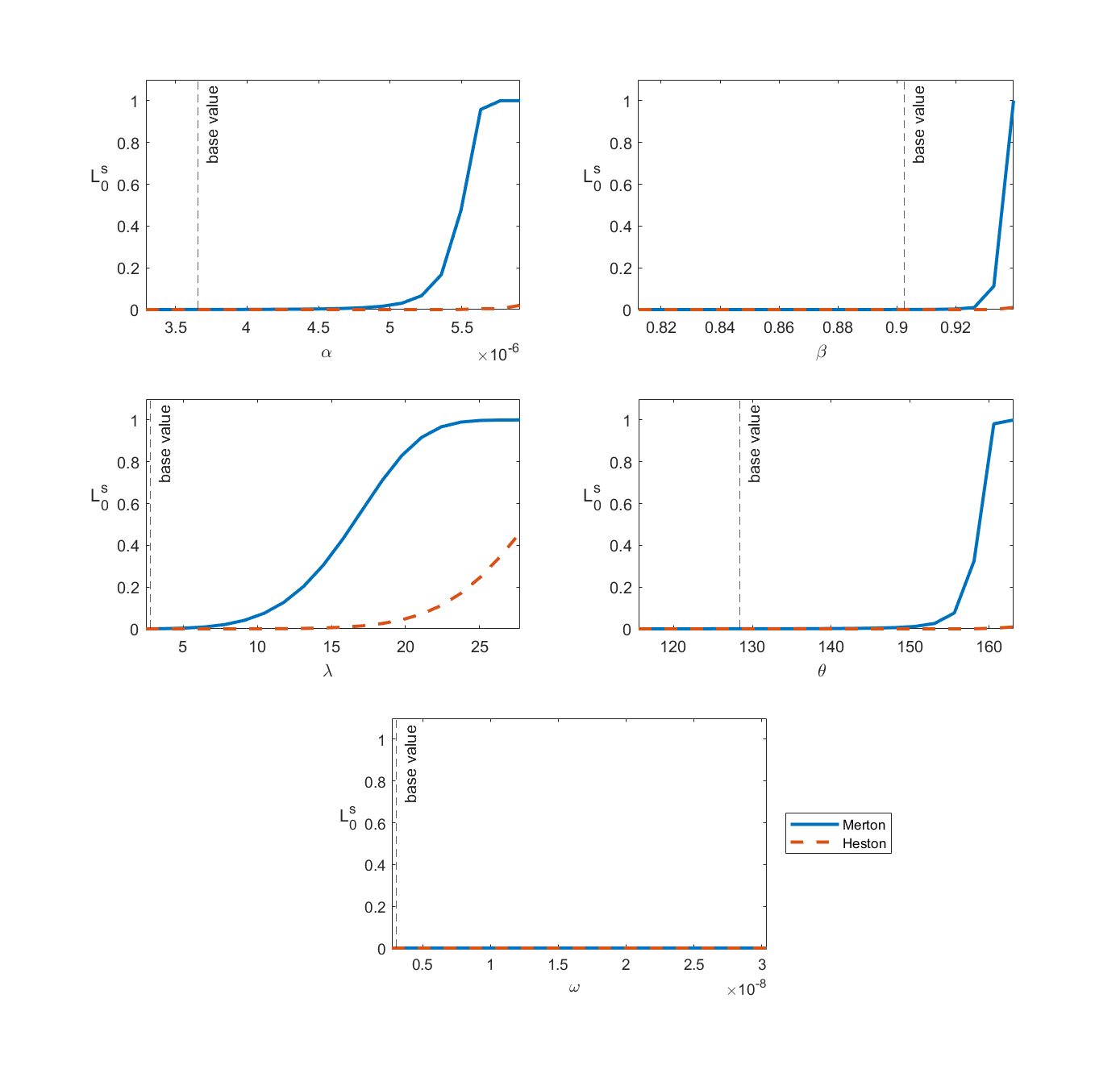}
\end{center}
\par
\noindent {\footnotesize This figure plots the WEL induced by following
Merton's and Heston's strategy for different parameter configurations.}
\end{figure}

In Figure \ref{fig:WELvsParams} we report the WEL when exploring other
reasonable values for each parameter while keeping all other parameters
unchanged. When varying the parameters we ensure that the stationarity
condition as well as the technical condition $1-2\alpha E_{t,T}^{s}>0$ from
Section \ref{sec:subopt} are satisfied. Other parameter regions than the
ones shown in the plots do not reveal any additional insights.

Recall, the WEL is calculated via the formula in Lemma \ref{lem WEL}. For
the purpose of these numerical examples we set $h_{t+1} = h$ using the
original parameters from \cite{Christoffersen.2006}. This gives us an
annualized volatility at time $t+1$ of $0.1578$ which is not modified along
with the model parameters. The reason for this is that if we recalculate $%
h_{t+1}$ using the varied parameters, $h_{t+1}$ would inflate as $\alpha$, $%
\beta$ or $\theta$ approach values that only just fulfill the stationarity
condition. This would cause the WEL to explode, i.e. $exp()$ in Lemma \ref%
{lem WEL} goes to $-\infty$ due to $\gamma<0$. However, we are interested in
the WEL that is induced by the difference in the strategies i.e. $%
D_{t,T}^{s}-D_{t,T}$ and/or $E_{t,T}^{s}-E_{t,T}^{\ast}$ and not by the
inflation of $h_{t+1}$. Keeping $h_{t+1}$ unchanged implies that Figure \ref%
{fig:WELvsParams} can answer the following question. If we observe an
annualized variance of $0.1578$ today, what would be the impact of changing
one model parameter on the WEL?

We observe that WEL rises as the values of $\alpha$, $\beta$, $\theta$ or $%
\lambda$ increase. For the first three parameters, the loss inferred by
following the Heston strategy is negligible, the one inferred by Merton's
solution on the other hand rises significantly. This behavior is reasonable
because if these parameters rise, then the variance becomes more random
hence creating a larger separation to Merton. The Merton solution completely
disregards the fact that variance is random in the HN-GARCH setting.
Explicitly modeling the variance gives additional information about how the
variance moves that can be exploited to achieve a higher expected utility or
more precisely the highest possible expected utility. $\omega$ has almost no
impact on WEL whatsoever because the optimal strategy is not sensitive to
changes in $\omega$ as demonstrated in Section \ref{sec:solSens}. $\lambda$
i.e. the equity risk premium has a significant impact on WEL. Recall from
Figure \ref{fig:sensParam}, that $\lambda$ already had a significant impact
on the overall strategy. However, from representation (\ref{opt pi Merton
plus 1}) one can see that it not only plays a role in the myopic but also in
the non-myopic part of the strategy. As its influence on the myopic part is
the same as for Merton's strategy one can follow from Figure \ref%
{fig:WELvsParams} that the non-myopic part becomes more and more important
for larger values of $\lambda$. Economically, the WEL induced by following
Merton's strategy can be explained as follows. $\lambda$ is the risk premium
per unit of variance. Merton assumes a constant variance that means the
premium ($\lambda h_{t}$) he can collect is always the same. If one knows
how variance moves over time, one can better incorporate how the premium ($%
\lambda h_{t}$) moves over time and thus collect this premium better. When
looking at the Heston strategy, one can see that $\lambda$ also has the
biggest impact on WEL. As the only difference between the Heston strategy
and ours is the size of one period this has to be the reason for the WEL
induced by following the Heston strategy. A possible economic interpretation
is that Heston over- or underestimates the opportunity to exploit the risk
premium by assuming the risky asset moves continuously while it actually
does only move at certain points in time. Furthermore, it is likely that WEL
significantly increases when using larger rebalancing frequencies.

In summary, WEL induced by following either Merton's or the Heston solution
instead of ours is rather small within this parameter set. However, there
are reasonable regions of parameters where WEL increases to a considerable
level.

\section{Conclusion\label{sec:CONCLUSION}}

This paper develops a closed-form solution for a portfolio
optimization problem where the investor maximizes a CRRA utility from
terminal wealth assuming that the variance of the spot asset follows the
HN-GARCH model. To avoid the possibility of a negative wealth we employ an
approximation for the wealth process.

In our two-asset setting, we find that the optimal portfolio process is
independent of the development of the risky asset and non-myopic i.e. has a
time-dependent component. In the limit where the length of one period goes
to zero our solution converges to the one under the Heston stochastic
volatility model under some conditions. Further we obtain a recursive
representation for the conditional m.g.f. of the optimal (and any other)
wealth process facilitating risk management and pricing.

Finally, using empirically relevant parameter estimates, we conduct a
numerical analysis of our optimal strategy. Firstly, we find that the
approximation of the wealth process has only minor impact. Secondly, our
strategy is quite robust against changes in parameter values. Thirdly, we
visualize the connection between our solution, the one under the Heston
model and the one under the Merton model. Our strategy contains both of the other
strategies as a special case i.e. the case $\Delta \longrightarrow 0$
and the case $\Delta = T$. And lastly, we investigate the WEL induced by
following the Heston and the Merton strategy for different parameter
settings. We report that, for a daily trading scenario, the optimal solution under the Heston model shows a good performance while the optimal
solution under the Merton model shows significant losses.

Overall, this paper provides a quick method of finding utility-optimal
portfolios under an advanced time-series model which is easy to
implement and takes into account important stylized facts.

\bibliographystyle{apalike}
\bibliography{References}

\newpage
\appendix

\section{Proofs}

\noindent\textbf{Proof of Theorem \ref{myTheorem 1}.\label{proofSpec 1}} We
prove the statements (\ref{ValueOpt_pert 1})-(\ref{OptAllocation_periodt 1})
from Theorem \ref{myTheorem 1} by induction.\newline
\textit{Base case} We start by optimizing the last period first i.e. from $T - 1$ to $T$ and use this as the base case. Let us solve problem (\ref{P(W) 1}) backward using Bellman's principle. The
first step reads%
\begin{align*}
\Phi_{T-1} (W_{T-1},h_T)=\underset{\pi _{T-1}}{\max }\mathbb{E}_{T-1}\left[
\frac{\exp \left\{ \gamma W_{T}\right\} }{\gamma }\right]
\end{align*}
where from (\ref{log wealth 1}) we have $W_{T} = W_{T-1}+r+\left( \pi _{T-1}-\pi _{T-1}^{2}\right) \frac{1}{2} h_{T}+\pi _{T-1}\left( \lambda h_{T}+\sqrt{h_{T}}z_{T}\right)$.\\
From \cite{Heston.2000} we know the one-period m.g.f.
of the log stock under the HN-GARCH model $\Psi _{X_{t}}^{\left( t-1\right) }
$ is
\begin{align*}
\Psi _{X_{T}}^{\left( T-1\right) }\left( u\right) &=\mathbb{E}
_{T-1}[e^{uX_{T}}]=\exp \left( uX_{T-1}+A_{T-1,T}+B_{T-1,T}h_{T}\right) \\
A_{T-1,T} &=ur,\quad B_{T-1,T}=u\lambda +\dfrac{1}{2}u^{2}.
\end{align*}
Using this we can produce the m.g.f. for log wealth as%
\begin{align}
\Psi _{W_{t}}^{\left( t-1\right) }\left( u\right) &=\mathbb{E}_{t-1}\left[
\exp \left\{ uW_{t}\right\} \right]  \notag \\
&=\Psi _{X_{t}}^{\left( t-1\right) }\left( u\pi _{t-1}\right) \times \exp
\left\{ u\left( W_{t-1}-\pi _{t-1}X_{t-1}+\left( \pi _{t-1}-\pi
_{t-1}^{2}\right) \frac{1}{2}h_{t}+\left( 1-\pi _{t-1}\right) r\right)
\right\}  \notag \\
&=\exp \left\{ uW_{t-1}+ur+\left( \pi _{t-1}u\lambda +\pi _{t-1}^{2}\dfrac{1%
}{2}u^{2}+\left( \pi _{t-1}-\pi _{t-1}^{2}\right) u\frac{1}{2}\right)
h_{t}\right\}  \label{mgf W interim}
\end{align}
Note this is the exponential of a polynomial of order two on $\pi _{t-1}$,
which will be used to maximize on $\pi _{t-1}$. With this we can rewrite the
problem as
\begin{align}
&\Phi_{T-1} (W_{T-1},h_T)=\underset{\pi _{T-1}}{\max }\mathbb{E}_{T-1}\left[
\frac{\exp \left\{ \gamma W_{T}\right\} }{\gamma }\right] =\underset{\pi
_{T-1}}{\max }\frac{1}{\gamma }\Psi _{W_{T}}^{\left( T-1\right) }\left(
\gamma \right)  \notag \\
\overset{(\ref{mgf W interim})}{&}{=}\underset{\pi _{T-1}}{\max }\frac{1}{%
\gamma }\exp \left\{ \gamma W_{T-1}+\gamma r +\left( \pi _{T-1}\gamma
\lambda +\pi _{T-1}^{2}\dfrac{1}{2}\gamma ^{2}+\gamma \left( \pi _{T-1}-\pi
_{T-1}^{2}\right) \frac{1}{2}\right) h_{T}\right\}  \notag \\
&=\underset{\pi _{T-1}}{\max }\frac{1}{\gamma }\exp \left\{ D_{T-1,T}+\gamma
W_{T-1}+E_{T-1,T}(\pi _{T-1})h_{T}\right\}  \label{T1 goal fct general 1}
\end{align}
where $D_{T-1,T}=\gamma r, \quad E_{T-1,T}(\pi _{T-1}) =\pi _{T-1}\left( \lambda +%
\frac{1}{2}\right) \gamma +\pi _{T-1}^{2}\dfrac{1}{2}\left( \gamma
^{2}-\gamma \right)$. Before carrying on, note that in (\ref{T1 goal fct general 1}) only $%
E_{T-1,T}(\pi_{T-1})$ depends on $\pi_{T-1}$ and that the exponential
function is strictly monotonously increasing in $E_{T-1,T}(\pi_{T-1})$.
Therefore, it is sufficient to only optimize $E_{T-1,T}(\pi_{T-1})$.
Further, note that the sign of the second derivative of the overall
optimization problem is only depending on the signs of $\gamma$ and $%
E_{T-1,T}^{\prime \prime }$. Therefore, it is also sufficient to calculate $%
E_{T-1,T}^{\prime \prime }$. With this we can proceed to solve the optimization problem in (\ref{T1 goal
fct general 1}) by taking the first derivative of $E_{T-1,T}(\pi_{T-1})$
w.r.t. $\pi_{T-1}$ and set it equal to zero,
\begin{align}
& \left( \lambda +\frac{1}{2}\right) \gamma +\pi _{T-1}^{\ast }\left( \gamma
^{2}-\gamma \right) =0 \Leftrightarrow \pi _{T-1}^{\ast }=\frac{\lambda +\frac{1}{2}}{1-\gamma }.
\end{align}
To check whether this is a minimum or a maximum we calculate the second
derivative of $E_{T-1,T}(\pi _{T-1})$ as $\gamma^{2}-\gamma>0, \quad \gamma<0.$ This shows that the optimum from above is maximizing if $\gamma<0$. However,
one can not be sure that the solution is minimizing for $\gamma>0$.
Substituting $\pi^{\ast}_{T-1}$ into equation (\ref{T1 goal fct general 1})
leads to
\begin{align}
\Phi_{T-1} (W_{T-1},h_T) &=\frac{1}{\gamma}\exp \left\{ D_{T-1,T}+\gamma
W_{T-1}+E_{T-1,T}^{\ast }h_{T}\right\}  \label{Opt Wealth T-1 1} \\
&=\frac{1}{\gamma}\exp \left\{ \gamma r+\gamma W_{T-1}+\frac{\gamma(\lambda +%
\frac{1}{2})^{2}}{2(1-\gamma)}h_{T}\right\}  \label{Phi interim}
\end{align}
where $E_{T-1,T}^{\ast }=E_{T-1,T}(\pi _{T-1}^{\ast })=(\lambda +\frac{1}{2}%
)^{2}\left(\frac{\gamma}{1-\gamma}+\frac{1}{2}\frac{1}{(1-\gamma)^{2}}%
(\gamma^{2}-\gamma)\right)= \frac{\gamma(\lambda +\frac{1}{2})^{2}}{%
2(1-\gamma)}$. Hence, $\Phi_{T-1}\in \mathbb{M}$. We can rewrite the terminal condition (\ref%
{Terminal Condition 1}) as
\begin{align}
\Phi_T (W_{T},h_{T+1})=\Phi_T (W_{T}) = \frac{\exp \left\{ \gamma W_{T}\right\} }{\gamma } \in \mathbb{M}.
\end{align}
This in combination with equation (\ref{Opt Wealth T-1 1}) imposes terminal
conditions on $D_{t,T}$ and $E_{t,T}^{\ast}$: $D_{T,T} = 0$, and $E_{T,T}^{\ast} = 0$.

\textit{Inductive step} Next we will perform the inductive step by assuming that the statements hold for a particular $t+1$ and showing that in this case they also hold for $t$. Moving one step backward from $t+1$ to $t$ by applying Bellman's principle
like before we get after plugging in the definitions of $W_{t+1}$ and $%
h_{t+2}$ and some algebraic manipulation:
\begin{align*}
&\Phi_t (W_{t},h_{t+1})=\underset{\pi _{t}}{\max }\mathbb{E}_{t}\left[ \Phi_{t+1}
(W_{t+1},h_{t+2})\right] =\underset{\pi _{t}}{\max }\mathbb{E}_{t}\left[\frac{1}{%
\gamma} \exp \left\{ D_{t+1,T}+\gamma
W_{t+1}+E_{t+1,T}^{\ast}h_{t+2}\right\} \right] \\
&= \underset{\pi _{t}}{\max }\left\{
\begin{array}{c}
\frac{1}{\gamma}\exp (D_{t+1,T}+\gamma \left(W_{t}+r+\pi_{t}\lambda
h_{t+1}+\left( \pi _{t}-\pi _{t}^{2}\right) \frac{1}{2}h_{t+1}\right) \\
+ E_{t+1,T}^{\ast}\left(\omega+\beta h_{t+1}+\alpha \theta^{2}
h_{t+1}\right)) \\
\times \mathbb{E}_{t}\left[ \exp \left\{\left(\gamma
\pi_{t}-2E_{t+1,T}^{\ast}\alpha \theta\right)\sqrt{h_{t+1}}%
z_{t+1}+E_{t+1,T}^{\ast}\alpha z_{t+1}^{2}\right\} \right]%
\end{array}
\right\}
\end{align*}
A useful result for a standard normal variable $z$ is that, like in \cite{Heston.2000},
\begin{align}
\mathbb{E}[\exp (aX^{2}+bX)]&=\dfrac{1}{\sqrt{(1-2a)}}\exp \left( \dfrac{
b^{2}}{2(1-2a)}\right).  \label{lemma HN z property}
\end{align}
With this we reduce the expectation to
\begin{align*}
&\mathbb{E}_{t}\left[ \exp \left\{\left(\gamma
\pi_{t}-2E_{t+1,T}^{\ast}\alpha\theta\right)\sqrt{h_{t+1}}%
z_{t+1}+E_{t+1,T}^{\ast}\alpha z_{t+1}^{2}\right\} \right] \\
&= \dfrac{1}{\sqrt{1-2E_{t+1,T}^{\ast}\alpha}}\exp \left( \dfrac{\left(
\gamma \pi_{t}-2E_{t+1,T}^{\ast}\alpha\theta \right)^{2}}{%
2(1-2E_{t+1,T}^{\ast}\alpha)}h_{t+1}\right).
\end{align*}
Here we need $1-2 \alpha E_{t,T}^{\ast}>0$ for the solution to be well
defined, again. Plugging this in and grouping yields
\begin{align}
\Phi_t (W_{t},h_{t+1}) &=\underset{\pi _{t}}{\max }\dfrac{1}{\gamma \sqrt{%
1-2E_{t+1,T}^{\ast}\alpha}}  \notag \\
&\times \exp \left\{
\begin{array}{c}
D_{t+1,T}+\gamma \left(W_{t}+r\right) + E_{t+1,T}^{\ast}\omega \\
+\left(\gamma \left(\lambda + \frac{1}{2} \right)\pi_{t}-\gamma\frac{1}{2}%
\pi _{t}^{2} + E_{t+1,T}^{\ast}\left(\beta+\alpha \theta^{2}\right)+\dfrac{%
\left( \gamma \pi_{t}-2E_{t+1,T}^{\ast}\alpha\theta \right) ^{2}}{%
2(1-2E_{t+1,T}^{\ast}\alpha)}\right) h_{t+1}%
\end{array}
\right\}  \notag \\
&=\underset{\pi _{t}}{\max }\frac{1}{\gamma}\exp \left\{ D_{t,T}+\gamma
W_{t}+E_{t,T}(\pi _{t})h_{t+1}\right\}  \label{t goal fct 1}
\end{align}
where
\begin{align}
D_{t,T} &=E_{t+1,T}^{\ast}\omega+\gamma r+D_{t+1,T}-\log \left(\sqrt{%
(1-2\alpha E_{t+1,T}^{\ast })}\right) \\
E_{t,T}(\pi _{t}) &=\left( \beta +\alpha \theta ^{2}\right)E_{t+1,T}^{\ast }
+\dfrac{\left( \gamma \pi _{t}-2\theta \alpha E_{t+1,T}^{\ast }\right) ^{2}}{%
2(1-2\alpha E_{t+1,T}^{\ast })} + \gamma \left(\left(\lambda + \frac{1}{2}
\right)\pi_{t}-\frac{1}{2}\pi_{t}^{2}\right).
\end{align}
Solving the optimization problem in (\ref{t goal fct 1}) yields
\begin{align*}
\frac{1}{2(1-2 \alpha E_{t+1,T}^{\ast})}\left(\gamma ^{2}2\pi_{t}^{\ast}-4
\gamma E_{t+1,T}^{\ast}\alpha\theta\right)-\gamma\pi_{t}^{\ast}+\gamma
\left(\lambda + \frac{1}{2}\right)=0
\end{align*}
\begin{align}
\Leftrightarrow \pi_{t}^{\ast} &= \left(\frac{2\gamma
E_{t+1,T}^{\ast}\alpha\theta- \gamma\left(\lambda + \frac{1}{2}\right)(1-2
\alpha E_{t+1,T}^{\ast})}{1-2 \alpha E_{t+1,T}^{\ast}} \right)\left(\frac{%
1-2 \alpha E_{t+1,T}^{\ast}}{\gamma^{2}-\gamma(1-2 \alpha E_{t+1,T}^{\ast})}%
\right)  \notag \\
&= \frac{\lambda + \frac{1}{2}}{(1-2 \alpha E_{t+1,T}^{\ast})-\gamma} -
\frac{\left(\theta+(\lambda+\frac{1}{2})\right) 2\alpha E_{t+1,T}^{\ast}}{%
(1-2 \alpha E_{t+1,T}^{\ast}) - \gamma}.  \label{opt pi Merton plus}
\end{align}
To check whether this is a minimum or a maximum we calculate the second
derivative of $E_{t,T}(\pi _{t})$ as $
\frac{1}{\underbrace{1-2\alpha E_{t+1,T}^{\ast}}_{>0}}\underbrace{\gamma ^{2}%
}_{>0}\underbrace{-\gamma}_{>0}>0, \quad \gamma<0$. Therefore, one can see that the assumption $\gamma<0$ is sufficient but not
necessary for any general $t$. If $\gamma>0$, we can not be sure that it is
minimizing. Plugging $\pi_{t}^{\ast}$ into equation (\ref{t goal fct 1})
yields $\Phi_t (W_{t},h_{t+1}) =\frac{1}{\gamma}\exp \left\{ D_{t,T}+\gamma
W_{t}+E_{t,T}^{\ast}h_{t+1}\right\}$ where in terms of $E^{\ast}_{t+1,T}$
\begin{align}
E_{t,T}^{\ast } = \left( \beta +\alpha \theta ^{2}\right)E_{t+1,T}^{\ast } +%
\dfrac{\left( \gamma \pi_{t}^{\ast}-2\theta \alpha E_{t+1,T}^{\ast}\right)
^{2}}{2(1-2\alpha E_{t+1,T}^{\ast})}+\gamma \left((\lambda + \frac{1}{2}%
)\pi_{t}^{\ast}-\frac{1}{2}(\pi_{t}^{\ast})^{2}\right).
\end{align}
Hence, $\Phi_{t}\in \mathbb{M}$. Starting from the induction hypothesis that the
statement holds for $t+1$ we have shown that in this case it also holds for $%
t$.\newline
%\textit{\underline{Proof of statements (\ref{ValueOpt_pert 1})-(\ref%
%{OptAllocation_periodt 1}): Conclusion}}\newline
%To sum up the results of the induction proof, since both the base case and
%the inductive step have been proved as true, by mathematical induction we
%have shown that the statements in Theorem \ref{myTheorem 1} hold for every $t
%$. \medskip

\noindent\textit{\underline{Proof of (\ref{W_opt_1}) as representation of $%
W_{t}^{\ast}$.}} We conjecture that the analytical expression of $W_{t}^{\ast}$ is given by Equation \ref{W_opt_1}.
%\begin{align}
%W^{\ast}_{t} &= w_{0}+ \sum_{k=1}^{t} \left((\lambda +\frac{1}{2})\pi
%_{t-k}^{\ast}-\frac{1}{2}(\pi_{t-k}^{\ast})^{2}\right) h_{t+1-k} +
%\sum_{k=1}^{t} \left(\pi_{t-k}^{\ast} \sqrt{h_{t+1-k}}z_{t+1-k}\right) +rt.
%\label{W_series 1 apdx}
%\end{align}
We will prove it by induction. The \textit{Base case} is straight forward to see that if we plug $t=1$ into the conjecture, we recover (\ref{W_rec_opt 1}) which is the definition of $W_{t}^{\ast}$. Thus, the statement holds for the base case.\newline
\textit{Inductive step:} The induction hypothesis is that Equation \ref{W_opt_1} holds for a particular $t$. It follows by adding $\left((\lambda +%
\frac{1}{2})\pi _{t}^{\ast}-\frac{1}{2}(\pi _{t}^{\ast})^{2}\right)h_{t+1} +
\pi _{t}^{\ast}\sqrt{h_{t+1}}z_{t+1}+r$ on both sides of the equation that
\begin{align*}
&W^{\ast}_{t} +\left((\lambda +\frac{1}{2})\pi _{t}^{\ast}-\frac{1}{2}(\pi
_{t}^{\ast})^{2}\right)h_{t+1} + \pi _{t}^{\ast}\sqrt{h_{t+1}}z_{t+1}+r \\
&= w_{0}+ \sum_{k=1}^{t} \left((\lambda +\frac{1}{2})\pi _{t-k}^{\ast}-\frac{%
1}{2}(\pi_{t-k}^{\ast})^{2}\right) h_{t+1-k} + \sum_{k=1}^{t}
\left(\pi_{t-k}^{\ast} \sqrt{h_{t+1-k}}z_{t+1-k}\right) +rt \\
&+\left((\lambda +\frac{1}{2})\pi _{t}^{\ast}-\frac{1}{2}(\pi
_{t}^{\ast})^{2}\right)h_{t+1} + \pi _{t}^{\ast}\sqrt{h_{t+1}}z_{t+1}+r.
\end{align*}
By (\ref{W_rec_opt 1}) the left-hand side of this equation is equal to $%
W_{t+1}^{\ast}$. The right-hand side can be written as
\begin{align*}
&w_{0}+ \sum_{k=1}^{t+1} \left((\lambda +\frac{1}{2})\pi _{t+1-k}^{\ast}-%
\frac{1}{2}(\pi_{t+1-k}^{\ast})^{2}\right) h_{(t+1)+1-k} + \sum_{k=1}^{t+1}
\left(\pi_{t+1-k}^{\ast} \sqrt{h_{(t+1)+1-k}}z_{(t+1)+1-k}\right) +r(t+1).
\end{align*}
Equating both sides, we deduce the statement also holds true for $t+1$.\newline
%\textit{Conclusion:}\newline
%Since both the base case and the inductive step have been proved as true, by
%mathematical induction the representation holds for every $t$. \medskip

\newpage

\section{Complementary material}

%\title{Closed-form portfolio optimization under garch models. \tnoteref{t1}}
%%\tnotetext[t1]{\color{red}{The authors thank \dots}}
%
%\author{Marcos Escobar-Anel\corref{cor1}}
%\ead{marcos.escobar@uwo.ca}
%\address{Department of Statistical and Actuarial Sciences, University of Western Ontario, London, ON, Canada, N6A5B7}
%\author{Maximilian Gollart}
%\ead{maximilian.gollart@tum.de}
%\author{Rudi Zagst}
%\ead{zagst@tum.de}
%\address{Department of Mathematics, Technical University of Munich, Munich, Germany}

\bigskip

\noindent\textbf{Proof of Lemma \ref{lem:mult period exp h}}
Using (\ref{HN GARCH}) we can write
\begin{align}
\mathbb{E}[h_{t}] &= \alpha + \omega +(\beta + \alpha \theta^{2})\mathbb{E}%
[h_{t-1}].  \label{E_{0}[h_t] recursive}
\end{align}
This we can use as a starting point and plug in the analogue expressions for
$\mathbb{E}[h_{t-1}], \mathbb{E}[h_{t-2}], \dots, \mathbb{E}[h_{0}]$. From
this we conjecture that the analytical representation is given by
\begin{align}
\mathbb{E}[h_{t}] &= \left(\alpha + \omega
\right)\sum_{i=0}^{t-1}\left((\beta + \alpha \theta^{2})^{i}\right) + (\beta
+ \alpha \theta^{2})^{t}h_{0}.  \label{exp h series}
\end{align}
Now that we have a guess for the analytical expression of $\mathbb{E}[h_{t}]$%
, we will prove it by induction next.

\noindent\textit{Base case:}\newline
We use $t=1$ as a base case which is true by (\ref{E_{0}[h_t] recursive}).

\noindent\textit{Inductive step:}\newline
The induction hypothesis is that (\ref{exp h series}) holds for a particular
$t$. It follows by multiplying by $\left(\beta + \alpha \theta^{2}\right)$
and adding $(\alpha+\omega)$ on both sides of the equation that
\begin{align*}
\mathbb{E}[h_{t}]\left(\beta + \alpha \theta^{2}\right)+\alpha+\omega= \left[%
\left(\alpha + \omega \right)\sum_{i=0}^{t-1}\left((\beta + \alpha
\theta^{2})^{i}\right) + (\beta + \alpha \theta^{2})^{t}h_{0}\right]%
\left(\beta + \alpha \theta^{2}\right)+\alpha+\omega.
\end{align*}
By (\ref{E_{0}[h_t] recursive}) the left-hand side of this equation is equal
to $\mathbb{E}[h_{t+1}]$. The right-hand side can be written as
\begin{align*}
\left(\alpha + \omega \right) \left[1+\sum_{i=1}^{t}\left((\beta + \alpha
\theta^{2})^{i}\right)\right] + (\beta + \alpha \theta^{2})^{t+1}h_{0} =
\left(\alpha + \omega \right) \sum_{i=0}^{t}\left((\beta + \alpha
\theta^{2})^{i}\right) + (\beta + \alpha \theta^{2})^{t+1}h_{0}.
\end{align*}
Equating both sides, we deduce the statement for $t+1$, establishing the
inductive step.

\noindent\textit{Conclusion:}\newline
Since both the base case and the inductive step have been proved as true, by
mathematical induction statement (\ref{exp h series}) holds for every $t$.

Up to this point we have proven that statement (\ref{exp h series}) is true.
By observing that the first part of this representation is a geometric
series we can get
\begin{align}
\mathbb{E}[h_{t}] = \left(\alpha + \omega \right) \left(\frac{1-\left(\beta
+\alpha \theta ^{2}\right)^{t}}{1-\left(\beta +\alpha \theta ^{2}\right)}%
\right) + (\beta + \alpha \theta^{2})^{t}h_{0}.  \label{E_{0}[h_t]}
\end{align}
\medskip

\noindent\textbf{Proof of Corollary \ref{myCorollary 1}. \label{proof mgfW 1}%
} The proof of the fact that the process is a GARCH model of the
Heston-Nandi type is straight forward. Knowing that $\left\{\pi_{t}^{\ast}%
\right\}_{t\in\left\{0,1,\dots, T\right\}}$ is deterministic one can see
that (\ref{W_rec_opt 1}) has the exact same form as the HN-GARCH model.%
\newline
\textit{\underline{Proof of conditional recursive m.g.f.}}\newline
\textit{Step 1:} We first compute the one-period moment generating function
of $W_{t+1}^{\ast}$,
\begin{align}
\mathbb{E}[e^{uW_{t+1}^{\ast}}\mid \mathcal{F}_{t}] \overset{(\ref{W_rec_opt
1})}{&}{=}\mathbb{E}[\exp \left(uW_{t}^{\ast} + u\left((\lambda +\frac{1}{2}%
)\pi _{t}^{\ast}-\frac{1}{2}(\pi _{t}^{\ast})^{2}\right)h_{t+1} + u\pi
_{t}^{\ast}\sqrt{h_{t+1}}z_{t+1} + ur\right) \mid \mathcal{F}_{t}]  \notag
\label{step 1 W 1} \\
& =\exp \left(uW_{t}^{\ast} + u\left((\lambda +\frac{1}{2})\pi _{t}^{\ast}-%
\frac{1}{2}(\pi _{t}^{\ast})^{2}\right)h_{t+1} + ur\right)\mathbb{E}[e^{u\pi
_{t}^{\ast}\sqrt{h_{t+1}}z_{t+1}}\mid \mathcal{F}_{t}]  \notag \\
\overset{(\ref{lemma HN z property})}{&}{=}\exp \left( uW_{t}^{\ast} +
u\left((\lambda +\frac{1}{2})\pi _{t}^{\ast}-\frac{1}{2}(\pi
_{t}^{\ast})^{2}\right)h_{t+1} + ur\right) \exp \left(0+\dfrac{1}{2}%
u^{2}\left(\pi_{t}^{\ast}\right)^{2}h_{t+1}\right)  \notag \\
& =\exp \left( uW_{t}^{\ast}+A_{t,t+1}+B_{t,t+1}h_{t+1}\right)
\end{align}%
with $A_{t,t+1}$ and $B_{t,t+1}$ given by
\begin{align}
A_{t,t+1}&=ur,\quad B_{t,t+1}=u\left((\lambda +\frac{1}{2})\pi _{t}^{\ast}-%
\frac{1}{2}(\pi _{t}^{\ast})^{2}\right) +\dfrac{1}{2}u^{2}\left(\pi_{t}^{%
\ast}\right)^{2}  \label{step 1 A1 B1 W 1}
\end{align}
\textit{Step 2:} We now compute the two-periods m.g.f of $W_{t+2}^{\ast}$
using the tower rule of expectations
\begin{align*}
\mathbb{E}[e^{uW_{t+2}^{\ast}}\mid \mathcal{F}_{t}]=\mathbb{E}[\mathbb{E}%
[e^{uW_{t+2}^{\ast}}\mid \mathcal{F}_{t+1}]\mid \mathcal{F}_{t}].
\end{align*}%
Let us move outward with the inner expectations. Using equations (\ref{step
1 W 1}) and (\ref{step 1 A1 B1 W 1}) we get
\begin{align*}
\mathbb{E}[e^{uW_{t+2}^{\ast}}& \mid \mathcal{F}_{t+1}]=\exp \left(
uW_{t+1}^{\ast}+A_{t+1,t+2}+B_{t+1,t+2}h_{t+2}\right) \\
A_{t+1,t+2}& =ur,\quad B_{t+1,t+2}=u\left((\lambda +\frac{1}{2})\pi
_{t+1}^{\ast}-\frac{1}{2}(\pi _{t+1}^{\ast})^{2}\right) +\dfrac{1}{2}%
u^{2}\left(\pi_{t+1}^{\ast}\right)^{2}.
\end{align*}%
Next,
\begin{align*}
& \mathbb{E}[e^{uW_{t+2}^{\ast}}\mid \mathcal{F}_{t}] \\
& =\mathbb{E}[\mathbb{E}[e^{uW_{t+2}^{\ast}}\mid \mathcal{F}_{t+1}]\mid
\mathcal{F}_{t}] \\
& =\mathbb{E}[\exp \left(
uW_{t+1}^{\ast}+A_{t+1,t+2}+B_{t+1,t+2}h_{t+2}\right) \mid \mathcal{F}_{t}]
\\
& =\mathbb{E}\left[ \exp \left(
\begin{array}{c}
u\left(W_{t}^{\ast} + \left((\lambda +\frac{1}{2})\pi _{t}^{\ast}-\frac{1}{2}%
(\pi _{t}^{\ast})^{2}\right)h_{t+1} + \pi _{t}^{\ast}\sqrt{h_{t+1}}%
z_{t+1}+r\right) +A_{t+1,t+2} \\
+B_{t+1,t+2}\left( \omega +\beta h_{t+1}+\alpha \left( z_{t+1}-\theta \sqrt{
h_{t+1}}\right) ^{2}\right)%
\end{array}%
\right) \mid \mathcal{F}_{t}\right] \\
& =\exp \left(
\begin{array}{c}
uW_{t}^{\ast}+u\left((\lambda +\frac{1}{2})\pi _{t}^{\ast}-\frac{1}{2}(\pi
_{t}^{\ast})^{2}\right) h_{t+1}+ur+A_{t+1,t+2} \\
+\omega B_{t+1,t+2}+B_{t+1,t+2}\beta h_{t+1} +B_{t+1,t+2}\alpha \theta
^{2}h_{t+1}%
\end{array}%
\right) \\
& \times \mathbb{E}\left[ \exp \left(B_{t+1,t+2}\alpha
z_{t+1}^{2}+z_{t+1}\left(u\pi_{t}^{\ast}\sqrt{h_{t+1}}-2\theta \sqrt{ h_{t+1}%
}B_{t+1,t+2}\alpha\right)\right) \mid \mathcal{F}_{t}\right]
\end{align*}%
To proceed, we use(\ref{lemma HN z property}) to calculate the expectation.
This gives
\begin{align}
& \mathbb{E}[e^{uW_{t+2}^{\ast}}\mid \mathcal{F}_{t}]  \notag \\
& =\exp \left(
\begin{array}{c}
uW_{t}^{\ast}+u\left((\lambda +\frac{1}{2})\pi _{t}^{\ast}-\frac{1}{2}(\pi
_{t}^{\ast})^{2}\right) h_{t+1}+ur+A_{t+1,t+2} \\
+\omega B_{t+1,t+2}+B_{t+1,t+2}\beta h_{t+1} +B_{t+1,t+2}\alpha \theta
^{2}h_{t+1}%
\end{array}%
\right)  \notag \\
& \times \exp \left( -\dfrac{1}{2}\log (1-2B_{t+1,t+2}\alpha )+\dfrac{\left(
u\pi_{t}^{\ast}\sqrt{h_{t+1}}-2\theta \sqrt{ h_{t+1}}B_{t+1,t+2}\alpha
\right) ^{2}}{ 2(1-2B_{t+1,t+2}\alpha )}\right)  \notag \\
& =\exp \left( uW_{t}^{\ast}+A_{t,t+2}+B_{t,t+2}h_{t+1}\right)
\label{step 2 W 1}
\end{align}%
with $A_{t,t+2},B_{t,t+2}$ given by
\begin{align}
A_{t,t+2}& =ur+A_{t+1,t+2}+\omega B_{t+1,t+2}-\dfrac{1}{2}\log
(1-2B_{t+1,t+2}\alpha )  \label{step 2 A2 W 1} \\
B_{t,t+2}& = u\left((\lambda +\frac{1}{2})\pi _{t}^{\ast}-\frac{1}{2}(\pi
_{t}^{\ast})^{2}\right)+B_{t+1,t+2}\left(\beta+\alpha \theta ^{2}\right)+%
\dfrac{\left( u\pi_{t}^{\ast}-2\theta B_{t+1,t+2}\alpha \right) ^{2}}{%
2(1-2B_{t+1,t+2}\alpha )}  \label{step 2 B2 W 1} \\
A_{t+1,t+2}& =ur,\quad \\
B_{t+1,t+2}& =u\left((\lambda +\frac{1}{2})\pi _{t+1}^{\ast}-\frac{1}{2}(\pi
_{t+1}^{\ast})^{2}\right) +\dfrac{1}{2}u^{2}\left(\pi_{t+1}^{\ast}\right)^{2}
\end{align}%
So we obtain the m.g.f of $W_{t+2}^{\ast}$ as
\begin{equation*}
\mathbb{E}[e^{u W_{t+2}^{\ast}}\mid \mathcal{F}_{t}]=\exp \left(
uW_{t}^{\ast}+A_{t,t+2}+B_{t,t+2}h_{t+1}\right).
\end{equation*}%
Observe that $A_{t,t+2},B_{t,t+2}$ are functions of $A_{t+1,t+2},B_{t+1,t+2}$
and $u$. Continuing recursively we can find the m.g.f of $W_{T}^{\ast}$ as
\begin{align*}
\mathbb{E}_{t}[e^{u W_{T}^{\ast}}]&=\exp \left(
uW_{t}^{\ast}+A_{t,T}+B_{t,T}h_{t+1}\right) \\
A_{t,T}&=A_{t+1,T}+ur+B_{t+1,T}\omega -\dfrac{1}{2}\log (1-2\alpha
B_{t+1,T}),\quad t=0,\dots ,T-1 \\
B_{t,T}&=u\left((\lambda +\frac{1}{2})\pi _{t}^{\ast}-\frac{1}{2}(\pi
_{t}^{\ast})^{2}\right) +\left(\beta+\alpha \theta ^{2}\right) B_{t+1,T}+%
\dfrac{(u\pi_{t}^{\ast}-2\alpha \theta B_{t+1,T})^{2}}{2(1-2\alpha
B_{t+1,T}\,)},\quad t=0,\dots ,T-1  \notag \\
A_{T,T}&=B_{T,T}=0.
\end{align*}

\noindent\textbf{Proof of Corollary \ref{cor:multiperiodExp}.}
	This follows by virtue of Theorem \ref{myTheorem 1} combined with the formula for the multi-period expectation of $h_{t}$ from Lemma \ref{lem:mult period exp h} yielding
	\begin{align*}
	&\mathbb{E}_{0}[W_{t}^{\ast}] = w_{0}+rt\\
	&+ \sum_{k=1}^{t} \left[
	\left((\lambda +\frac{1}{2})\pi _{t-k}^{\ast}-\frac{1}{2}(\pi
	_{t-k}^{\ast})^{2}\right)
	\left(\left(\alpha + \omega \right) \left(\frac{1-\left(\beta +\alpha	\theta ^{2}\right)^{(t+1-k)}}{1-\left(\beta +\alpha\theta ^{2}\right)}\right) + (\beta + \alpha \theta^{2})^{(t+1-k)}h_{0}\right)\right] \\
	&+ \sum_{k=1}^{t} \left(\pi_{t-k}^{\ast} \mathbb{E}_{0}[\sqrt{h_{t+1-k}}z_{t+1-k}]\right).
	\end{align*}
From (\ref{HN GARCH}) we can see that $h_t$ depends on the path of $z$ until $t-1$ but is independent of $z_{t}$. Thus we get
	\begin{align*}
	&\mathbb{E}_{0}[W_{t}^{\ast}] = w_{0}+rt\\ &+\sum_{k=1}^{t} \left[
	\left((\lambda +\frac{1}{2})\pi _{t-k}^{\ast}-\frac{1}{2}(\pi
	_{t-k}^{\ast})^{2}\right)
	\left(\left(\alpha + \omega \right) \left(\frac{1-\left(\beta +\alpha	\theta ^{2}\right)^{(t+1-k)}}{1-\left(\beta +\alpha\theta ^{2}\right)}\right) + (\beta + \alpha \theta^{2})^{(t+1-k)}h_{0}\right)\right].
	\end{align*}

\noindent\textbf{Proof of Lemma \ref{lem WEL}.}
	As for Lemma \ref{myProp1} and Theorem \ref{myTheorem 1}, we have
	\begin{align}
	\Phi_t^{s} (\log(V_{t}),h_{t+1})&=\frac{1}{\gamma }\exp \left\{ D_{t,T}^{s}+\gamma \log(V_{t})+E_{t,T}^{s}h_{t+1}\right\} \\
	\Phi_t (\log(V_{t}(1-L_{t}^{s})),h_{t+1})&=\frac{1}{\gamma }\exp \left\{ D_{t,T}+\gamma \log(V_{t}(1-L_{t}^{s}))+E_{t,T}^{\ast
	}h_{t+1}\right\}.
	\end{align}
	Plugging this into (\ref{WEL_impl}) yields
	\begin{align}
	&\frac{1}{\gamma }\exp \left\{ D_{t,T}+\gamma \log(V_{t}(1-L_{t}^{s}))+E_{t,T}^{\ast
	}h_{t+1}\right\} = \frac{1}{\gamma }\exp \left\{ D_{t,T}^{s}+\gamma \log(V_{t})+E_{t,T}^{s}h_{t+1}\right\} \notag \\
	\Leftrightarrow &\gamma \log(V_{t})+\gamma \log(1-L_{t}^{s}) = \gamma \log(V_{t})+\left(E_{t,T}^{s}-E_{t,T}^{\ast
	}\right)h_{t+1}+\left(D_{t,T}^{s}-D_{t,T}\right) \notag \\
	\Leftrightarrow &L_{t}^{s} = 1 - \exp\left\{\frac{1}{\gamma}\left(\left(D_{t,T}^{s}-D_{t,T}\right) + \left(E_{t,T}^{s}-E_{t,T}^{\ast
	}\right)h_{t+1}\right)\right\} \notag.
	\end{align}

\noindent\textbf{Proof of Proposition \ref{myProp}.}
To present the limit of the wealth process (\ref{W	delta v}), we consider the parameter assumptions from \cite{Badescu.2019},
\begin{align}
\alpha (\Delta )&=\alpha \Delta \text{ (i.e. } \alpha_h(\Delta)=\alpha \Delta^{2}),\quad \theta (\Delta )=\dfrac{\theta }{%
	\sqrt{\Delta }}\text{ (i.e. } \theta_h(\Delta)=\frac{\theta}{\Delta}),\quad \lambda (\Delta )=\lambda.  \label{param assumptions W}
\end{align}%
The conditional first and second moments of the log-wealth processes (\ref{W	delta v}) are
given by
\begin{align}
\mathbb{E}_{t-\Delta}[W_{t}-W_{t-\Delta}] &= r(\Delta) + \pi
_{t-\Delta}\lambda v_{t}\Delta+\left( \pi _{t-\Delta}-\pi
_{t-\Delta}^{2}\right) \frac{1}{2}v_{t}\Delta  \label{moments1 1} \\
\mathbb{V}ar_{t-\Delta}[W_{t}-W_{t-\Delta}] &=\pi _{t-\Delta}^{2}v_{t}\Delta
\label{W var}
\end{align}
\begin{align}
\mathbb{C}ov_{t-\Delta}[v_{t+\Delta}-v_{t}, W_{t}-W_{t-\Delta}] &= \mathbb{C}%
ov_{t-\Delta}\left[
\begin{array}{c}
\omega(\Delta) +\beta(\Delta) v_{t} \\
+\alpha(\Delta) (z_{t}-\theta(\Delta) \sqrt{v_{t}})^{2} - v_{t}, \pi
_{t-\Delta}\sqrt{v_{t}}\sqrt{\Delta}z_{t}%
\end{array}%
\right]  \notag \\
&- \mathbb{C}ov_{t-\Delta}[\alpha(\Delta) 2z_{t}\theta(\Delta) \sqrt{v_{t}},
\pi _{t-\Delta}\sqrt{v_{t}}\sqrt{\Delta}z_{t}]  \notag \\
&= - 2\alpha(\Delta)\theta(\Delta) \pi _{t-\Delta}v_{t}\sqrt{\Delta}  \notag
\\
&= - 2\alpha\Delta\theta \pi _{t-\Delta}v_{t}  \label{W cov} \\
Corr_{t-\Delta}\left[v_{t+\Delta}-v_{t}, W_{t}-W_{t-\Delta}\right] &= \frac{%
	\mathbb{C}ov_{t-\Delta}[v_{t+\Delta}-v_{t}, W_{t}-W_{t-\Delta}]}{\sqrt{%
		\mathbb{V}ar_{t-\Delta}[W_{t}-W_{t-\Delta}]}\sqrt{\mathbb{V}%
		ar_{t-\Delta}[v_{t+\Delta}-v_{t}]}}  \notag \\
&= \frac{- 2\frac{\theta}{\sqrt{\Delta}} \sqrt{v_{t}}}{\sqrt{2 +4\frac{%
			\theta^{2}}{\Delta} v_{t}}}.  \label{moments1 -1}
\end{align}
To obtain the moments of the continuous-time process we calculate the limits
of the moments of the discrete-time process from equations (\ref{moments1 1}%
)-(\ref{moments1 -1}) per time step $\Delta$ as $\Delta \longrightarrow0$.
This reads
\begin{align}
\underset{\Delta \rightarrow 0}{\lim }\frac{\mathbb{E}_{t-%
		\Delta}[W_{t}-W_{t-\Delta}]}{\Delta} &= \underset{\Delta\rightarrow 0}{\lim}%
\frac{r(\Delta) + \pi _{t-\Delta}\lambda v_{t}\Delta+\left( \pi
	_{t-\Delta}-\pi _{t-\Delta}^{2}\right) \frac{1}{2}v_{t}\Delta}{\Delta}
\notag \\
&=r + \pi _{t}\lambda v_{t}+\left( \pi _{t}-\pi _{t}^{2}\right) \frac{1}{2}%
v_{t} \\
\underset{\Delta \rightarrow 0}{\lim }\frac{\mathbb{V}ar_{t-%
		\Delta}[W_{t}-W_{t-\Delta}]}{\Delta} &= \pi _{t}^{2}v_{t} \\
\underset{\Delta \rightarrow 0}{\lim }\frac{\mathbb{C}ov_{t-\Delta}[v_{t+%
		\Delta}-v_{t}, W_{t}-W_{t-\Delta}]}{\Delta} &= - 2\alpha\theta \pi _{t}v_{t}
\\
\underset{\Delta \rightarrow 0}{\lim }\frac{Corr_{t-\Delta}[v_{t+%
		\Delta}-v_{t}, W_{t}-W_{t-\Delta}]}{\Delta} &= -sign(\theta).
\end{align}
With this we can derive the diffusion limit and arrive at
\begin{align}
dW_{t} &= \left( r + \pi _{t}\lambda v_{t}+\left( \pi _{t}-\pi _{t}^{2}\right)
\frac{1}{2}v_{t}\right) dt+\pi _{t}\sqrt{v_{t}}dz^{S}_{t}
\end{align}
which is equivalent to the log-wealth process in the Heston model from
equation (\ref{W Heston}).

\end{document}